\documentclass[structabstract]{aa}
\usepackage{graphicx}
\usepackage{natbib}

\usepackage{txfonts}
\begin{document}

\title{Molecular chemistry and the missing mass problem in planetary nebulae}

\titlerunning{Molecular chemistry and the missing mass problem in PNe}

   \author{R. K. Kimura
          \inst{1}\thanks{E-mail: kimura@astro.iag.usp.br}
          \and
          R. Gruenwald\inst{1}
          \and
          I. Aleman\inst{1} \fnmsep \inst{2} \fnmsep \inst{3}
          }

\authorrunning{Kimura, Gruenwald and Aleman}

\institute{Instituto de Astronomia, Geof\'{i}sica e Ci\^encias
           Atmosf\'ericas, Universidade de S\~ao Paulo, 
           Rua do Mat\~ao 1226, 
           S\~ao Paulo,SP, 05508-090, Brazil
           \and
           Departamento de Astronomia, Instituto de F\'isica,
           Universidade Federal do Rio Grande do Sul, 
           Av. Bento Gon\c{c}alves 9500,
           Porto Alegre, RS, 91501-970, Brazil
           \and
           Jodrell Bank Centre for Astrophysics, 
           The Alan Turing Building, School of Physics and Astronomy, 
           The University of Manchester, Oxford Road, Manchester, 
           M13 9PL, UK\\
           }

   \date{Received November 10, 2011; accepted Month 00, 2011}

\abstract{Detections of molecular lines, mainly from H$_2$ and CO, reveal molecular material in planetary nebulae. Observations of a variety of molecules suggest that the molecular composition in these objects differs from that found in interstellar clouds or in circumstellar envelopes.  The success of the models, which are mostly devoted to explain molecular densities in specific planetary nebulae, is still partial, however.}
         {The present study aims at identifying the influence of stellar and nebular properties on the molecular composition of planetary nebulae by means of chemical models. A comparison of theoretical results with those derived from the observations may provide clues to the conditions that favor the presence of a particular molecule.}
         {A self-consistent photoionization numerical code was adapted to simulate cold molecular regions beyond the ionized zone. The code was used to obtain a grid of models and the resulting column densities are compared with those inferred from observations.}
         {Our models show that the inclusion of an incident flux of X-rays is required to explain the molecular composition derived for planetary nebulae. We also obtain a more accurate relation for the $N($CO$)/N($H$_2)$ ratio in these objects. Molecular masses obtained by previous works in the literature were then recalculated, showing that these masses can be underestimated by up to three orders of magnitude. We conclude that the problem of the missing mass in planetary nebulae can be solved by a more accurate calculation of the molecular mass.}
         {}


\maketitle

\section{Introduction} \label{sec:Introduction}

Molecular material is detected in many planetary nebulae (PNe), mainly by the identification of CO and H$_2$ lines. CO has been detected in about 100 PNe \citep{HugginsEtAl1996, HugginsEtAl2005}, while infrared emission of H$_2$ has been identified in more than 70 PNe \citep{HoraEtAl1999,SterlingAndDinerstein2008}. The observations, both in CO and H$_2$, indicate that these objects have in general a high or moderate N/O abundance ratio ($>0.3$) and bipolar morphology. The detections include young and evolved PNe, such as NGC 7027 and NGC 7293 (Helix Nebula), suggesting that molecules can survive over a significant fraction of the lifetime of a PN. 

The detection of molecules other than H$_2$ and CO is still restricted to a few PNe. Most molecules have been detected in the well-studied source NGC 7027 \citep[e. g.][]{LiuEtAl1996,LiuEtAl1997,HasegawaAndKwok2001} and in NGC 7293 \citep[][and references therein]{TenenbaumEtAl2009}. For other PNe, a leading work was done by \citet{BachillerEtAl1997}. They reported line observations of some molecules (CO, $^{13}$CO, CN, HCN, HNC, HCO$^+$, CS, SiO, SiC$_2$, and HC$_3$N) in a sample of seven objects at different stages of evolution. \citet{BachillerEtAl1997} showed that the molecular composition in PNe is different from that in stars in the asymptotic giant branch (AGB) and from that in proto-planetary nebulae (proto-PNe). As the star evolves from AGB to proto-PN and then to the PN phase, CS, SiO, SiC$_2$, and HC$_3$N densities decrease, such that these molecules are not detected in PNe. On the other hand, CN, HCN, HNC, and HCO$^+$ densities increase during the evolution. The HCO$^+$ emission is not detected or is extremely weak in progenitor objects such as IRC$+$10216 and CRL 2688, but it is relatively strong in the PNe phase. The study of \citet{BachillerEtAl1997} is complemented by observations of \citet{JosselinAndBachiller2003}, who reported observations of $^{13}$CO, CN, HCN, HNC, and HCO$^+$ in a sample of compact PNe, and by \citet{BellEtAl2007}, who carried out a search for CO$^+$ emission in eight objects, including proto-PNe and PNe.

\citet{Black1978} developed the first chemical model for PNe. He modeled the ionized gas and predicted the existence of diatomic molecules, such as H$_2$, H$_2^+$, HeH$^+$, OH, and CH$^+$, in the transition zone H$^+$/H$^0$. A steady-state chemical model for the neutral gas contained in globules of PNe was developed by \citet{HoweEtAl1994}. They predicted detectable quantities of CN, HCN, and HNC in a carbon-rich gas and concluded that their results agree qualitatively with the observations. For HCO$^+$, however, the resulting density is much lower than that obtained from the observational data.

Based on observations of the early-type PN NGC 7027 by the Infrared Space Telescope (ISO), \citet{YanEtAl1999} presented a thermal-chemical model for the neutral envelope of this object in a semi-infinite slab approximation. Their work indicates that the molecular densities obtained for NGC 7027 can be explained by a hot gas chemistry. \citet{HasegawaEtAl2000} developed a spherical symmetric, steady-state chemical model for the same PN, with emphasis on radiative transfer, but without a detailed thermal or dynamical treatment. Their results show that the molecular densities in NGC 7027 are a result of a combination of photochemistry and high gas temperature.

\citet{NattaAndHollenbach1998} developed time-dependent evolutionary models for the neutral envelope of PNe. Their models  include the H$_2$ chemistry and thermal balance. They studied the effect of shocks, far-ultraviolet radiation, and also soft X-ray emission from hot central stars.

\citet{AliEtAl2001} constructed a time-dependent chemical model to investigate the chemistry of clumpy neutral envelopes of three PNe: NGC 6781, M4-9, and NGC 7293. Compared with data inferred from the observations, the calculated fractional density of CN is too high by a factor of 2-3, while HCO$^+$ is less abundant by a factor of 5. Their models also predict too much CS and SiO.

Both \citet{NattaAndHollenbach1998} and \citet{AliEtAl2001} included X-ray emission from hot stars in their simulations. They showed that X-ray photons can affect the physical and chemical conditions in PNe. In particular, \citet{AliEtAl2001} reproduced the density ratios between CO, CN, and HCN by including X-ray effects in their simulations. However, observations show that the stellar emission is not the only possible source of X-rays in PNe. It is now well known that X-rays can have a diffuse origin \citep[][and references therein]{Kastner2007}. One possibility is that this emission is produced by a hot bubble, a 
rarefied gas of high temperature generated by wind interactions 
\citep[e.g.][]{MellemaEtAl1995}. X-ray emitting gas can also be in the form of collimated fast winds or jets \citep{SokerAndKastner2003}. 

In this paper we present the results of a self-consistent calculation of molecular concentrations in PNe, from the ionized region to the external neutral and cold gas. Our calculations span a wide range of physical parameters that characterize PNe. In order to compare our results with the observational data, we restrict our analysis to the molecules CO, HCO$^+$, CN, HCN, and HNC. 

The models are described in Sect. \ref{sec:models}. Model results are presented and discussed in Sect. \ref{sec:results}, 
while in Sect. \ref{sec:discussion_x} our results are compared with those derived from 
observations. A discussion about the determination of molecular masses 
as the missing mass problem is presented in Sect. \ref{sec:molecular_mass}. Conclusions are 
summarized in Sect. \ref{sec:conclusion}.

\section{Models} \label{sec:models}

In this section we present the numerical code used to obtain the models and the included chemical network. The assumed input parameters and the range of values adopted for the free parameters are discussed. 

\subsection{The numerical code}

The numerical code Aangaba \citep[][and references therein]{AlemanAndGruenwald2011} simulates the physical conditions in a nebula illuminated by an ionizing radiation source. The simulation starts at the inner border of the nebula and proceeds into the outward direction. In each position, the physical conditions (e.g., atomic, molecular, and electronic densities, and gas temperature), as well as continuum and line emissivities, are calculated. The thermal and chemical structures are mutually dependent, and depend on the incident radiation in the particular position in the nebula. Thus, the code performs iterative calculations to obtain the solution of the coupled equations. 

The outward-only approximation is adopted for the transfer of the primary and diffuse radiation fields. Geometric dilution and extinction of the radiation by gas and dust are taken into account. 

The density of the gas phase species (atoms, molecules, their respective ions, and electrons) are obtained at each position of the nebula under the chemical and ionization equilibrium hypotheses.  Twelve elements and their ions are included: H, He, C, N, O, Mg, Ne, Si, S, Ar, Cl, and Fe. 

The original code was improved for also dealing with neutral and cold regions. 
For this, a more extended set of molecules and chemical reactions was introduced (details in the following section) as well as cooling and heating mechanisms expected for a neutral gas. All temperature-dependent data, new or previously included in the code, were adapted to correctly treat a low-temperature gas. The references for data related to the gas temperature are the following: \citet{Aggarwal1983}, \citet{AggarwalEtAl1984, AggarwalEtAl1985}, \citet{HayesAndNussbaumer1984}, \citet{KeenanEtAl1986}, \citet{Martin1988}, \citet{Smits1991}, \citet{EkbergAndFeldman1993}, \citet{SaweyAndBerrington1993}, \citet{ZhangAndPradhan1995}, \citet{StoreyAndHummer1995}, \citet{QuinetEtAl1996}, \citet{RamsbottomEtAl1997}, \citet{AggarwalAndKeenan1999}, \citet{CleggEtAl1999}, \citet{VernerEtAl1999}, \citet{BrayEtAl2000}, \citet{GuptaAndMsezane2000}, \citet{Tayal2000}, \citet{WilsonAndBell2002}, and \citet[][and references therein]{FerlandEtAl2009}.

The gas temperature is calculated under the thermal equilibrium hypothesis, i. e. the total gain of energy per unit time and volume is balanced by the total loss of energy per unit time and volume. Several gas heating and cooling mechanisms due to atomic species, dust, 
molecules (H$_2$, CO, OH, and H$_2$O), as well as those driven by radiation, 
are included. Cosmic-ray heating is also included assuming the formalism of \citet{Yusef-ZadehEtAl2007} with an ionization rate of H$_2$ by cosmic-ray equal to $1.3 \times 10^{-17}$ s$^{-1}$ \citep{WoodallEtAl2007}. Dust is included as in Gruenwald et al. (in preparation). 
Heating and cooling mechanisms related to the H$_2$ molecule are described in \citet{AlemanAndGruenwald2011}. 

The level population of some molecules, required for calculating line intensities and the resulting gas cooling, is obtained assuming statistical equilibrium for H$_2$, CO, H$_2$O, and OH; local thermodynamic equilibrium is assumed for CN, HCN, HNC and HCO$^+$. For CO, H$_2$O, and OH the statistical equilibrium were carried out using the MOLPOP program developed by M. Elitzur (private comm.), which was coupled to the Aangaba code. The detailed treatment for the statistical equilibrium of H$_2$ can be found in \citet{AlemanAndGruenwald2011}. The line emissivities of CO, H$_2$O, and OH were calculated assuming the escape probability equation given by \citet{HollenbachAndMcKee1979}. The cooling due to these molecules is assumed to be the sum of the energy emitted by all allowed transitions between the energy levels included in the code.  Data for transition probabilities, collision coefficients, and energy levels were obtained from the LAMDA \citep{SchoierEtAl2005} and HITRAN 2005 \citep{RothmanEtAl2005} databases. 

\subsection{The chemical network} \label{sub:The_chemical_network} 

Our chemical model includes 95 molecules (see Table \ref{tab:chemical_chain_network}) besides the above mentioned atoms and their ions. The chemical network consists of 1693 reactions and is based on the UDFA 2006 catalog \citep{WoodallEtAl2007}. The adopted set of reactions is appropriate to describe the chemistry of the main species studied here according to the astrochemical literature \citep[e.g.][]{vanDishoeckAndBlack1989, SternbergAndDalgarno1995, BogerAndSternberg2005}. The rate coefficients are also taken from UDFA 2006 (including cosmic-ray ionization rates), except those related to photo-processes or to the H$_2$ network reactions. 

The rate coefficients of the photo-processes are obtained by integrating the cross-section over the energy distribution of the incident radiation. The cross-sections are those from the Huebner database  \citep{HuebnerEtAl1992} and \citet[][and references therein]{Dishoeck1987}. When a photo-process cross-section is not available, the value of the photoreaction coefficient ($\Gamma$) was estimated by a similarity criterion defined as

\begin{equation}
	\Gamma = \frac{\alpha}{\alpha_R}\Gamma_R,
	\label{eq:gamma_sem_sigma}
\end{equation}
\noindent where $\alpha$ (in s$^{-1}$) is the rate in the unshielded interstellar radiation field given in UDFA 2006 for each process. The $R$ subscript denotes ``reference'', indicating the value of the respective variable for the process used as reference. We chose the reference according to the value of the $\gamma$ parameter:

\begin{itemize}
	\item CH photodissociation for $1.0 > \gamma \leq 1.5$
	\item OH photodissociation for $1.5 > \gamma \leq 2.0$
	\item CO photodissociation for $\gamma > 2.0$
\end{itemize}
\noindent The $\gamma$ parameters relate the wavelength dependence of the photoprocess and are also given in UDFA 2006 for each molecule. 

The coefficients for the chemical H$_2$ network (which contains the H$_2$, H$_2^+$, H$_3^+$, and H$^-$ species) and details of the dust properties can be found in \citet{AlemanAndGruenwald2004, AlemanAndGruenwald2011}. The H$_2$ self-shielding is included following the formalism of \citet{BlackAndDishoeck1987}. The CO self-shielding and the CO cross-shielding by H$_2$ are estimated from the tabular values given by \citet{vanDishoeckAndBlack1988}.

\begin{table}[hbt]
\caption{Molecular species included in the chemical network.}
\label{tab:chemical_chain_network}
\begin{center}
\begin{tabular}
{ll|ll|l|l|l}
\hline
\multicolumn{2}{c}{2 atoms} & \multicolumn{2}{c}{3 atoms} & \multicolumn{1}{c}{4 atoms} & \multicolumn{1}{c}{5 atoms} &	\multicolumn{1}{c}{6 atoms} \\ 
\hline
	C$_2$	&	NH$^+$	&	C$_2$H	&	HCO	&	 CH$_3$	&	CH$_4$	&	CH$_5^+$	\\
	C$_2^+$	&	NO	&	C$_2$H$^+$	&	HCO$^+$	&	CH$_3^+$ &	CH$_4^+$	&		\\
	CH	&	NO$^+$	&	C$_2$N	&	HCS	&	C$_2$H$_2^+$	&	H$_3$CO$^+$	&		\\
  CH$^+$	&	NS	&	C$_2$N$^+$	&	HCS$^+$	&	C$_3$H$^+$	&	H$_3$CS$^+$	&		\\
	CN	&	NS$^+$	&	C$_3$	&	HNC	&	C$_2$NH$^+$	&	NH$_4^+$	&		\\
	CN$^+$	&	O$_2$	&	C$_3^+$	&	NH$_2$	&	H$_2$CO	&		&		\\
	CO	&	O$_2^+$	&	CH$_2$	&	NH$_2^+$  &	H$_2$CO$^+$	&		&		\\
	CO$^+$	&	OH	&	CH$_2^+$	&	O$_2$H$^+$	&	H$_2$CS	&		&		\\
  CS	&	OH$^+$	&	CNC$^+$	&	OCS	&	H$_2$CS$^+$	&		&		\\
	CS$^+$	&	SiC	&	CO$_2$	&	SiH$_2^+$	&	H$_2$NC$^+$	&		&		\\
	H$_2$	&	SiC$^+$	&	CO$_2^+$	&	SiOH$^+$	&	H$_3$O$^+$	&		&		\\
	H$_2^+$	&	SiH	&	H$_2$O	&	SO$_2$	&	H$_3$S$^+$	&		&		\\
	HeH$^+$	&	SiH$^+$	&	H$_2$O$^+$	&		&	HCNH$^+$	&		&		\\
	HS	&	SiO	&	H$_2$S	&		&	HCO$_2^+$	&		&		\\
	HS$^+$	&	SiO$^+$	&	H$_2$S$^+$	&		&	NH$_3$	&		&		\\
	N$_2$	&	SO	&	H$_3^+$	&		&	NH$_3^+$	&		&		\\
	N$_2^+$	&	SO$^+$	&	HCN	&		&	SiCH$_2^+$	&		&		\\
	NH	&		&	HCN$^+$	&		&		&		&		\\
\hline
\end{tabular}
\end{center}
\end{table}

\subsection{Parameters of the models}

We obtained a grid of theoretical models with input parameters in ranges typical of PNe to the understanding of the correlation between molecular concentrations and some properties of PNe. 

The primary energy sources are the central star and the X-ray emission produced in a hot bubble located in a cavity between the central star and the main PN shell. The central star is assumed to emit as a blackbody, defined by its temperature ($T_*$) and luminosity ($L_*$). The hot bubble emission is represented by a central source of X-rays, defined by the spectral distribution (0.3 - 2 keV) and by the integrated X-ray luminosity ($L_X$). The X-ray spectral distribution assumed is that derived for NGC 5315 by \citet{Kastner2007}.

The gas density and chemical composition, as well as the dust properties (density, composition and size) describe the nebula. The nebular gas density is represented by the total number of hydrogen nuclei ($n_H$). Most models were obtained with a homogeneous density. For testing the effect of density profiles in the molecular composition, models with different density profiles are also analyzed. The tested radial profiles are the following: a homogeneous distribution, a power law ($r^{-p}$, $1 \leq p \leq 4$), a distribution defined by equilibrium pressure, or a combination of them. 

The nebula is assumed to be spherically symmetric. The internal radius of the nebula is defined as $R_0 = 10^{15}$ cm. The exact value of $R_0$ does not affect the results presented in this paper for values up to 30 \% of the ionized radius. 

The adopted chemical composition, homogeneous throughout the nebula, corresponds to the mean values for PNe according to \citet{KingsburghAndBarlow1994}. The abundances of Mg, Si, Cl, and Fe are not provided by these authors. In these cases, we adopted the values given by \citet{StasinskaAndTylenda1986} to make a rough correction for grains depletion. Since these elements do not have much effect on the gas cooling or the resulting molecular densities, their exact proportion in the form of grains is not important. The abundances of C, N, and O may affect the resulting molecular composition. Therefore, we also analyzed models with different values for the abundance of these elements.

Dust is included as graphite spheres of radius $10^{-2} \mu$m. The dust-to-gas mass ratio ($M_d/M_g$) is assumed to be constant throughout the nebula.

The adopted ranges for the free parameters for the planetary nebulae models are given in Table \ref{tab:FreeParameters}, with the corresponding references. A standard model with a given set of input parameters was chosen, with the corresponding adopted values also listed in Table \ref{tab:FreeParameters}. To study the effects of each parameter individually, we varied one of the parameters within its typical range, while keeping the others fixed. Hereafter, unless otherwise noted, the parameters of the models are those of the standard model.

\begin{table}[hbt]
\caption{Model parameters: variation range, references and standard value.}
\label{tab:FreeParameters}
\begin{center}
\begin{tabular}
{lccc}
\hline
Parameter & Variation range & References & Standard value \\
\hline
$T_*$ (K) & $3 \times 10^4 - 3 \times 10^5$ & 1, 2 & $10^5$  \\
$L_*$ (L$_\odot$) & $10^2 - 1.2 \times 10^4$ & 1, 3 & $3 \times 10^3$ \\
$n_H$ (cm$^{-3}$)  & $10^{2}$ - $5 \times 10^{5}$ & 4, 5, 6 & $10^{5}$ \\
$C/H$ & $10^{-4}$ - $6 \times 10^{-4}$ & 7, 8 & $5.50 \times 10^{-4}$ \\
$N/H$ & $10^{-4}$ - $6 \times 10^{-4}$ & 7, 8 & $2.24 \times 10^{-4}$ \\
$O/H$ & $10^{-4}$ - $6 \times 10^{-4}$ & 7, 8 & $4.79 \times 10^{-4}$ \\
$M_{d}/M_{g}$ & $10^{-3}$ - $10^{-2}$ & 9 & $5 \times 10^{-3}$ \\
$L_{X}$ (erg s$^{-1}$) & $0; 10^{30}$ - $10^{32}$ & 10, 11 & $5 \times 10^{31}$ \\
\hline
\multicolumn{4}{p{9cm}}{\tiny {\bf References:} (1) \citet{BloeckerEtAl1995}; (2) \citet{Phillips2003}; (3) \citet{Phillips2005}; (4) \citet{LiuEtAl2001}; (5) \citet{StanghelliniAndKaler1989}; (6) \citet{KingsburghAndEnglish1992}; (7) \citet{KingsburghAndBarlow1994}; (8) \citet{MilanovaAndKholtygin2009}; (9) \citet{StasinskaAndSzcerba1999}; (10) \citet{Kastner2007}; (11) \citet{MellemaEtAl1995}.} 
\end{tabular}
\end{center}
\end{table}

In the following sections the discussion of the molecular composition is made mainly in terms of the column density of the CO molecule, $N($CO$)$, since this value is easily obtained from the observations, where the column density is the integral along the radius outward from the central star (both hemispheres are taken into account). Models are then obtained with the calculations stopping at different values of $N($CO$)$.

\section{Model results} \label{sec:results}

We outline here the thermal and chemical structures for the standard model. This first analysis provides a useful guide for the discussion about the molecular chemistry in models with different free parameters and for a comparison with the observations, which will both be presented in Sect. \ref{sec:discussion_x}. 

\subsection{Basic chemical structure of PNe} \label{}

Since the central star of PNe is an intense source of UV photons, ionized matter is present in any PNe. If there is enough mass to absorb the ionizing and dissociating photons, a given nebula can also have an outer region of neutral and cold matter (radiation- or ionization-bounded nebula). 

To simplify our discussion about the molecular chemistry in PNe we can identify some regions according to the predominant form of H, C, or O (ionic, neutral or molecular). We identified the following regions: H$^+$, H$^0$, H$_2$, and CO. As the name suggests, the main characteristics of the three first regions is the predominant form of hydrogen. The CO region is defined as the region where the CO molecule locks up all oxygen or carbon (the less abundant), i. e., where CO is fully associated. With this definition, the CO region corresponds to the external and coldest zones where hydrogen is molecular. We freely defined boundaries for these regions; from the inner to the more external regions, the boundaries are defined by the conditions $n($H$^+)=n($H$^0)$, $n($H$^0)=n($H$_2)$, and $n($CO$)=n($X$^0_{C,O})$, where $n($X$)$ is the volumetric density of the species X, and X$_{C,O}$ represents the less abundant element (C or O). Indeed, the transition from one region to another is smooth and the extent of the transition zones depends on the stellar and nebular characteristics. The transition regions may be relevant for the molecular production, since the coexistence of two or more forms of a given element may enhance or inhibit some molecular processes. For the present discussion, we distinguish three transition zones here: H$^+$/H$^0$, H$^0$/H$_2$, and C$^+$/C$^0$/CO.

The physical conditions in the H$^+$ region are regulated by energetic photons emitted by the hot central star ($T_*>3 \times 10^4$ K). In this region, electrons are produced mainly by the ionization of hydrogen and helium. The gas temperature is typically about $10^4$ K. In this harsh environment molecules can survive just in the transition zone toward the H$^0$ region.  In this transition zone (H$^+$/H$^0$) the coexistence of cations, anions, and the neutral atom of hydrogen in a warm temperature environment provides the formation of H$_2$ by alternative routes and not just by grain surface reactions \citep{AlemanAndGruenwald2004}.

In the H$^0$ region hydrogen is predominantly atomic. Here, the UV photons above 13.6 eV are absorbed by gas and dust in inner shells, while the flux that dissociates H$_2$ is intense. The physical conditions of the gas in this region are controlled by far-UV photons (6.0 - 13.6 eV) with high influence of X-rays photons (0.3 - 2.0 keV). The ejection of electrons from grains (photoelectric effect) is the main source of heating, balanced by cooling due to metals (mainly C, N, and O). The electrons are provided mainly by the ionization of H$^0$ ($n($H$^+)/n_H \sim 10^{-4}$), maintained by X-ray photons, and from the C$^0$ ionization by far-UV photons. Nitrogen and oxygen are predominantly in atomic form. This region extends up to the region where photons able to dissociate the molecular hydrogen are mostly absorbed. Under this condition, the hydrogen becomes predominantly molecular, characterizing the H$_2$ region.

In the H$_2$ region the heating and most chemical reactions are dominated by radiation. The cooling by CO becomes significant and dominates the total energy loss. Hydrogen is predominantly molecular (H$_2$) and has a strong influence on the chemical reactions, since H$_2$ facilitates the formation or hydrogenation of some molecules. Nitrogen and oxygen are still in atomic form. 

Depending on the spectral distribution and intensity of the incident spectrum, as well as on the nebular parameters, a nebula can be extended enough such that the radiation able to ionize C$^0$ or to dissociate CO is attenuated in the more external region. This is what we call the CO region, where the CO dissociation rate is very low. In this region the main source of energy gradually changes for increasing radius from the photoelectric effect to heating by cosmic rays. The cooling due to the CO molecule is the dominant mechanism of energy loss, and the electron density is determined by the ionization of C$^0$ or of heavy metals of low ionization potential (Mg, Fe), where C$^+$ abundance is low. Nitrogen is predominantly in molecular form (N$_2$), while oxygen is in CO form.  In this region the temperature reaches its lowest value ($T \approx 40$ K).

\subsection{Main molecular formation and destruction routes}

In this section we analyze the molecular distribution along the nebular radius. The molecular distribution is discussed in terms of the volumetric density (cm$^{-3}$), since this quantity provides local information about the molecular processes (and consequently the physical conditions) in different regions of the nebula. 

\begin{figure}[!htb]
\begin{center}
\includegraphics[width=1 \columnwidth,angle=0]{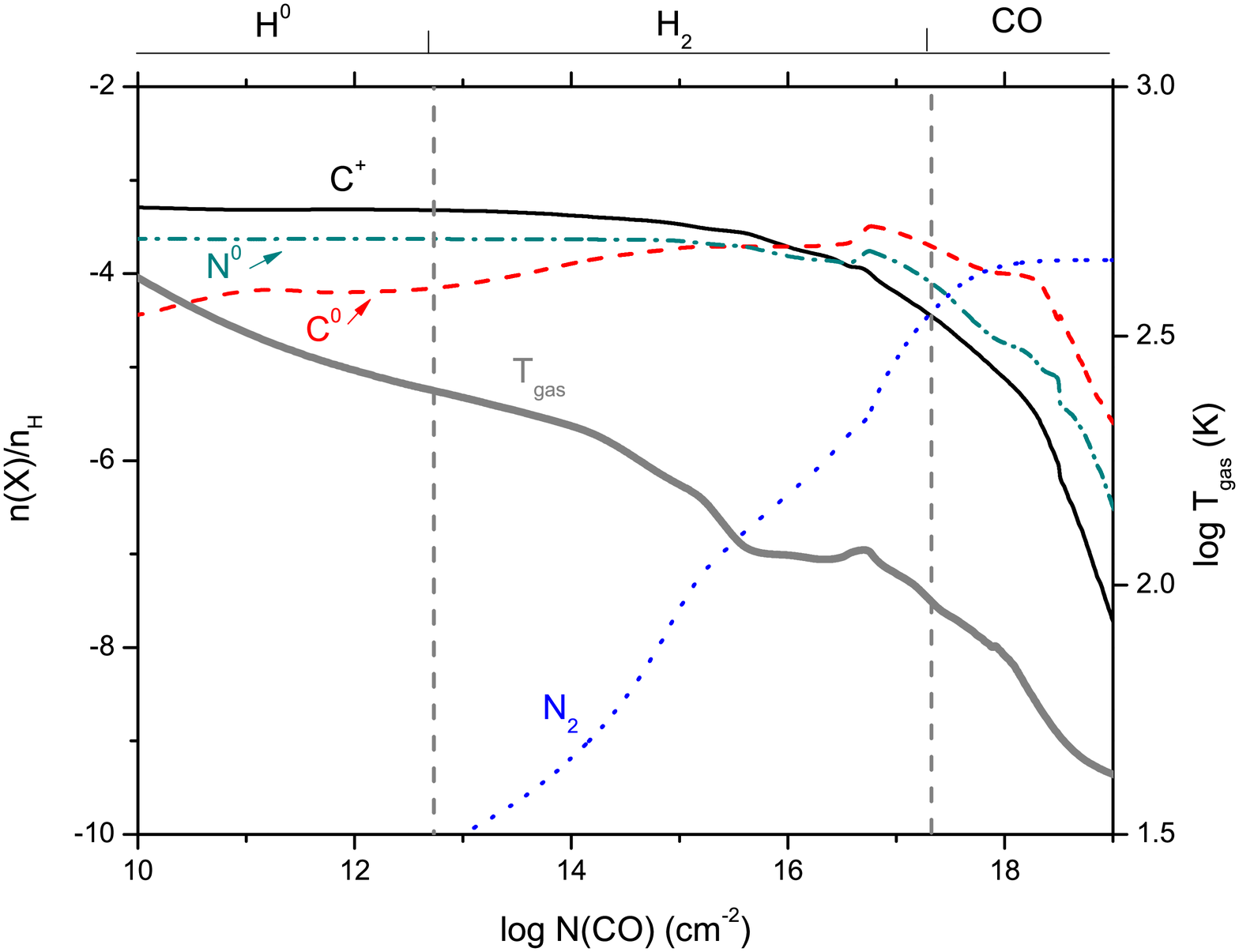}
\includegraphics[width=1 \columnwidth,angle=0]{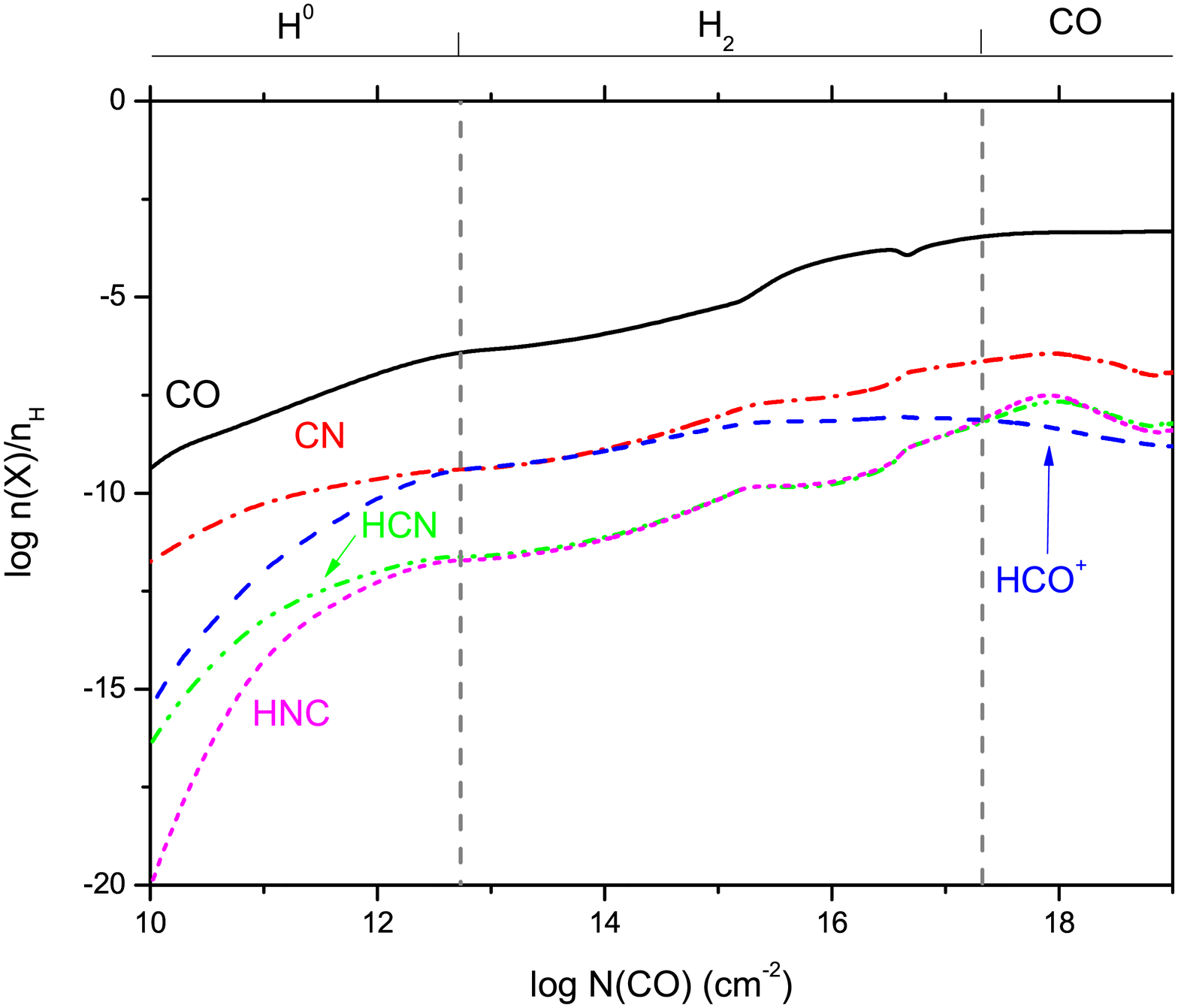}
\caption{Atomic and molecular abundances and gas temperature versus $N($CO$)$ for the standard model. The line at the top indicates the H$^0$, H$_2$ and CO regions. (A color version of this figure is available in the online journal).}
\label{fig:SM_abundances}
\end{center}
\end{figure}

\begin{figure}[!htb]
\begin{center}
\includegraphics[width=1 \columnwidth,angle=0]{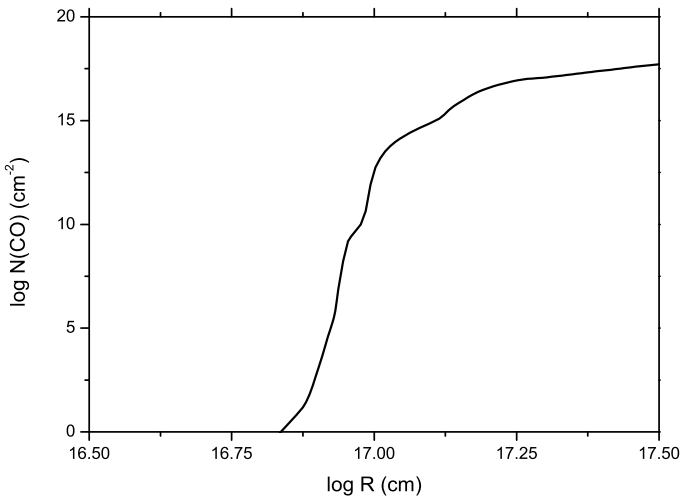}
\caption{Dependence of $N($CO$)$ on radius for the standard model.}
\label{fig:NCO_R}
\end{center}
\end{figure}

Figure \ref{fig:SM_abundances} shows the gas temperature and relative (to n$_H$) abundances of C$^0$, C$^+$, N$^0$, N$_2$, CO, HCO$^+$, CN, HCN, and HNC versus the CO column density, $N($CO$)$, for the standard model. The vertical lines in the figure indicate the boundaries of some of the previously defined regions. Following our definition, the lines corresponding to the standard model (from left to right) are $n($H$^0)=n($H$_2)$ and $n($CO$)=n($O$^0)$. The regions showed in the figure correspond, from left to right, to the H$^0$, H$_2$, and CO regions. The H$^+$ region is not included in Fig. \ref{fig:SM_abundances}, since molecular concentrations are negligible in this region. The dependence of $N($CO$)$ on the radius for the standard model is shown in Fig. \ref{fig:NCO_R}.

To understand the pattern presented by the density of the molecules along the nebula, shown in Fig. \ref{fig:SM_abundances}, we identify the main formation and destruction processes.

\begin{figure}[!htb]
\begin{center}
\includegraphics[width=1 \columnwidth,angle=0]{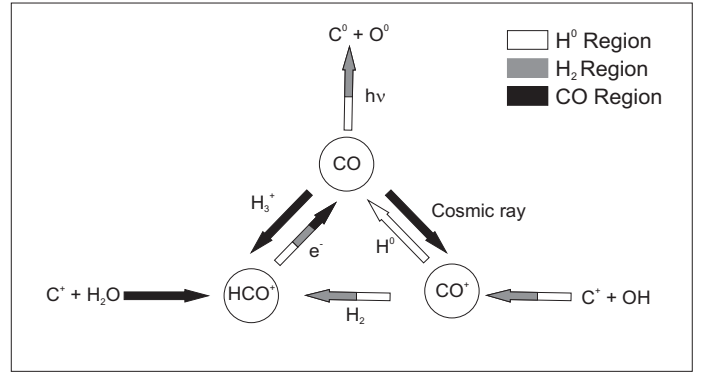}
\caption{Schematic diagram showing the chemical paths for formation and destruction of CO and HCO$^+$.}
\label{fig:diagrama_CO_COp_HCOp}
\end{center}
\end{figure}

Since the reaction chains of the molecules CO and HCO$^+$ are intrinsically coupled, their densities are mutually dependent. Figure \ref{fig:diagrama_CO_COp_HCOp} presents a schematic diagram of the main processes between CO and HCO$^+$. Details of the reactions summarized in the diagram are given below. 

One path for the formation of the radical HCO$^+$ is via the reaction

\[
\mathrm{CO} \stackrel{\mathrm{H}_3^+}{\rightarrow} \mathrm{HCO}^+.
\]
\noindent This reaction is efficient only in the CO region because it requires a high survival rate of H$_3^+$, which is expected to be very short-lived because of its rapid dissociative recombination in the presence of electrons, a consequence of an ultraviolet radiation field. An X-ray incident flux raises the density of H$_3^+$ by increasing its production rate; nonetheless, this is not the main source of HCO$^+$ if there is strong UV radiation. In the H$^0$ and H$_2$ regions where the UV radiation field is intense, the main formation route of HCO$^+$ is

\[
\mathrm{CO}^+ \stackrel{\mathrm{H}_2}{\rightarrow}  \mathrm{HCO}^+.
\]

The molecular ion CO$^+$ is formed mainly by the route

\[
\mathrm{O}^+ \stackrel{\mathrm{H}_2}{\rightarrow} \mathrm{OH}^+ \stackrel{\mathrm{H}_2}{\rightarrow} \mathrm{H}_2\mathrm{O}^+ \stackrel{\mathrm{H}_2}{\rightarrow} \mathrm{H}_3\mathrm{O}^+ \stackrel{\mathrm{e}^-}{\rightarrow} \mathrm{OH} \stackrel{\mathrm{C}^+}{\rightarrow} \mathrm{CO}^+.
\]

An X-ray incident flux is required for the formation of CO$^+$ molecules, since it contributes to the increase of the electron and O$^+$ densities. The most favorable condition for the CO$^+$ formation is found in the transition zone H$^0$/H$_2$, where there are enough H$_2$ molecules for the synthesis of OH, but their density is not high enough to efficiently convert CO$^+$ into HCO$^+$. In the CO region the main formation route of CO$^+$ is the ionization of CO by cosmic rays at a low production rate, leading to a very low density compared to the CO$^+$ formed in the H$^0$ and H$_2$ regions. 

CO is formed via  HCO$^+$ + e$^-$ $\rightarrow$ CO + H and
via  CO$^+$ + H$^0$ $\rightarrow$ CO + H$^+$ , although the
contribution from the latter is minor. The destruction of CO is dominated by radiation, not only by dissociation, but also by reactions involving He$^+$, which is a product of ionization by X-rays.

\begin{figure}[!htb]
\begin{center}
\includegraphics[width=1 \columnwidth,angle=0]{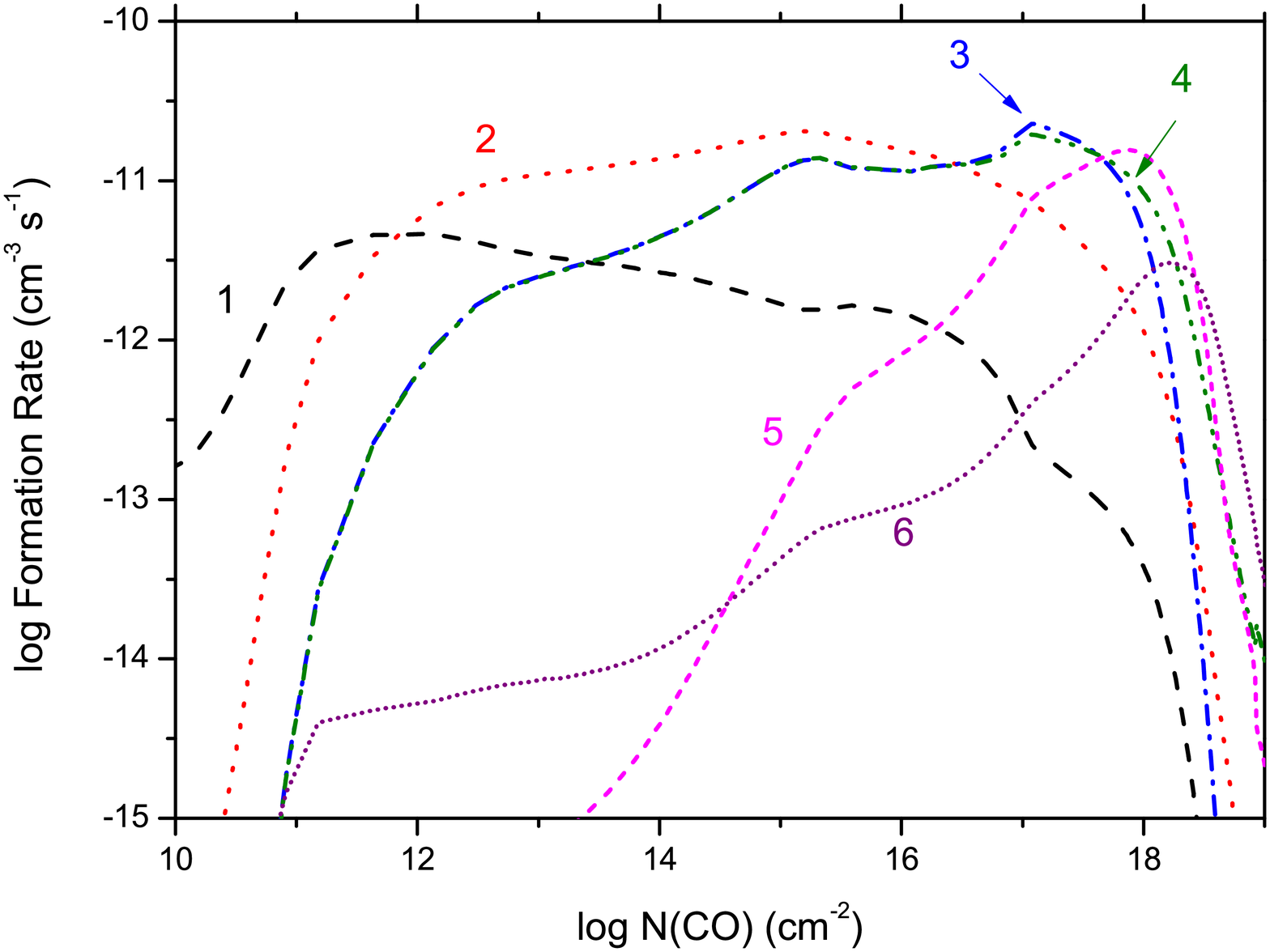}
\caption{Rates of the main reactions forming CN for the standard model. (1) H$^0$ + CN$^+$ $\rightarrow$ CN + H$^+$; (2) HCN$^+$ + e$^-$ $\rightarrow$ CN + H$^0$; (3) HCN + photon $\rightarrow$ CN + H$^0$; (4) HCNH$^+$ + e$^-$ $\rightarrow$ CN + H$_2$; (5) C$^0$ + NO $\rightarrow$ CN + O$^0$; (6) CH + N$^0$ $\rightarrow$ CN + H$^0$.}
\label{fig:CN_formation_rates}
\end{center}
\end{figure}

The scenario is more complicated for CN, since there are many reactions that can play the major role in forming the molecule across the nebula.  This can be verified in Fig. \ref{fig:CN_formation_rates}, which presents the formation rate of the main reactions leading to the CN molecule as a function of $N($CO$)$.  

For HCN, the main formation reaction in most of the nebula is

\[	
 \mathrm{HCNH}\mathbf{^+ \stackrel{\mathrm{e}^-}{\rightarrow}} \mathrm{HCN}.
\]

\noindent HCNH$^+$ can be produced by different routes triggered by ions generated by photons or by cosmic rays. The main route depends on the specific region and can include the reaction with activation barrier C$^+ +$ H$_2 \rightarrow$ CH$^+ + $H$^0$, which is processed in the H$^0$/H$^+$ transition zone (where T$_{gas} \gtrsim 10^3$ K). The charge exchange between H$^0$ and HCN$^+$ also contributes in the H$^0$ region and C$^0$ + NH$_2$ with some importance in the transition zone C$^+$/C$^0$/CO. Photodissociation is the main destruction reaction in the H$^0$ and H$_2$ regions, as in the inner zone of the CO region. When photons able to dissociate HCN are scarce, the destruction is dominated by collisions with products and subproducts of the ionization by cosmic rays.

The destruction of CN and HCN is dominated by photodissociation in most of the nebula. In outer zones of the CO region, where the photodissociation is not effective, the destruction is controlled by collisions with products and subproducts of ionization by cosmic rays, such as H$_3^+$ and HCO$^+$.

The chemistry of HNC can be described just by substituting HCN by HNC, with small differences in the rate coefficients of the reactions. 

\subsection{The effect of C/O and N/O ratios on the chemistry} \label{}

To discuss the C/O ratio we can distinguish two regimes for the molecular production: one in which the radiation controls the chemistry (in the H$^0$ and H$_2$ regions) and another in the CO region, where the main chemical processes are induced by cosmic-rays. In the H$^0$ and H$_2$ regions the molecular chemistry is not very sensitive to the C/O ratio, such that the molecular concentration in an oxygen-rich gas and in a carbon-rich gas is similar.

\begin{figure*}[!htb]
\begin{minipage}[t]{0.45\linewidth}
\centering
\includegraphics[width=1 \columnwidth,angle=0]{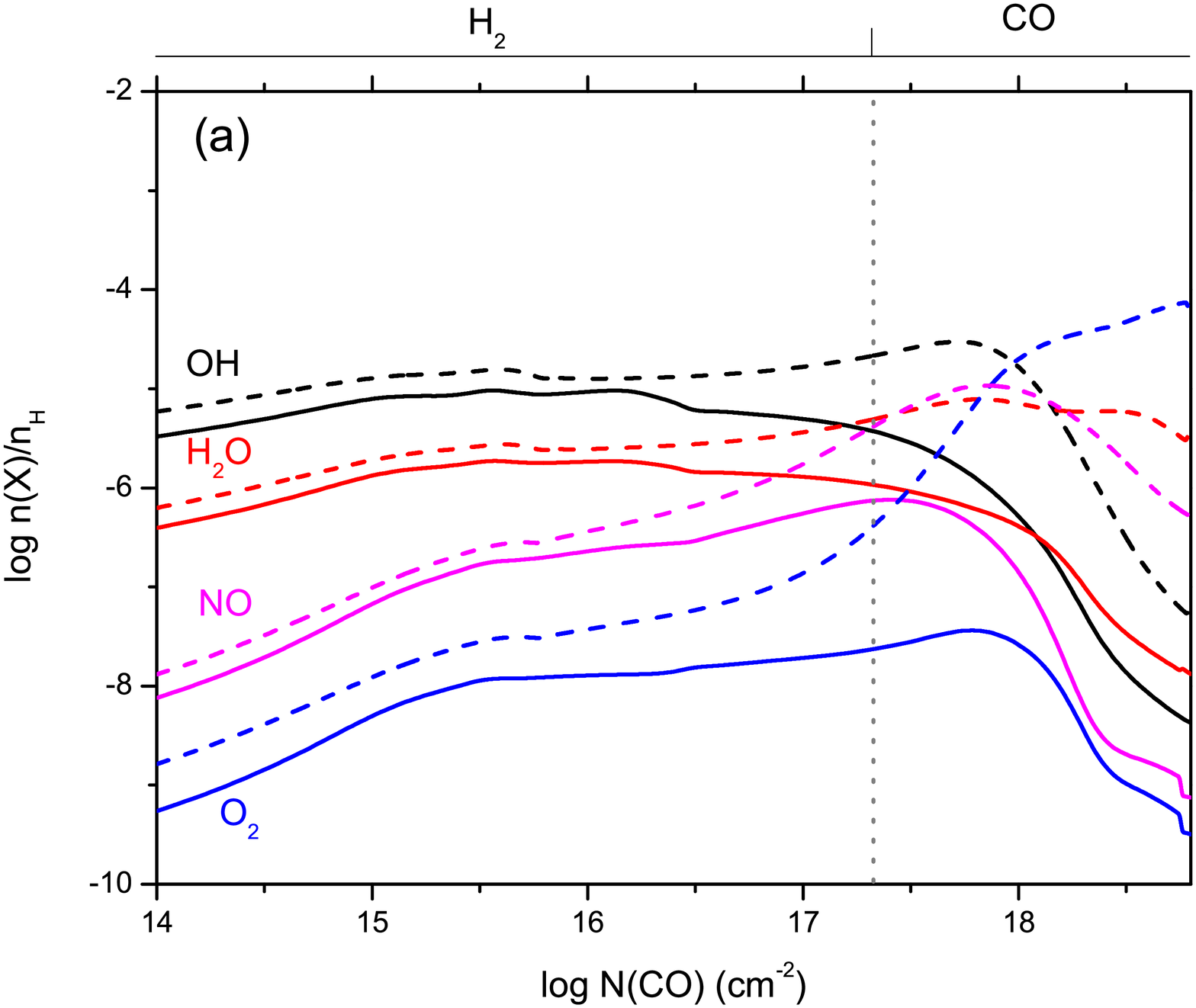}
\end{minipage}
\hspace{0.4cm}
\begin{minipage}[t]{0.45\linewidth}
\centering
\includegraphics[width=1 \columnwidth,angle=0]{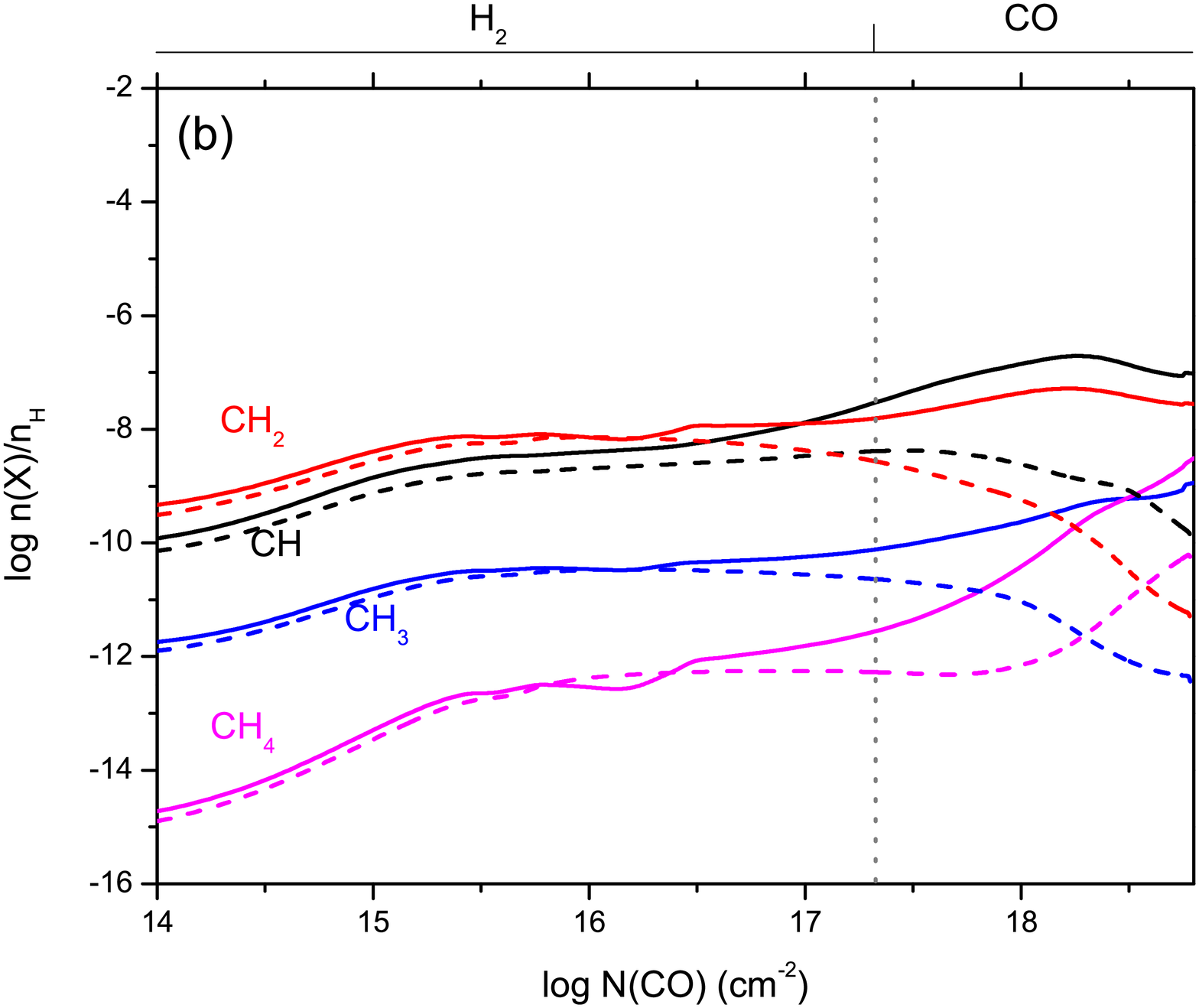}
\end{minipage}

\begin{minipage}[t]{0.45\linewidth}
\centering
\includegraphics[width=1 \columnwidth,angle=0]{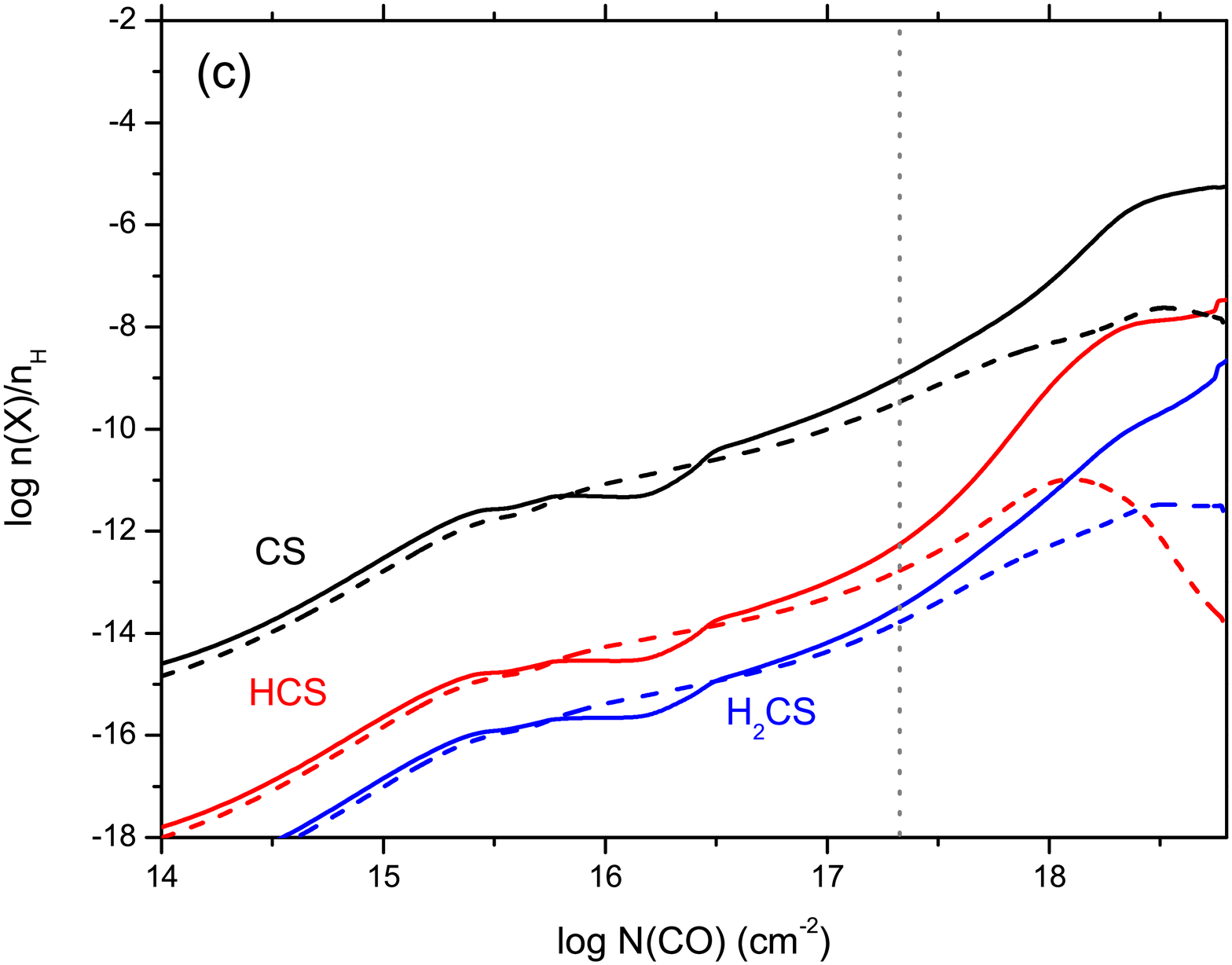}
\end{minipage}
\hspace{0.5cm}
\begin{minipage}[t]{0.45\linewidth}
\centering
\includegraphics[width=1 \columnwidth,angle=0]{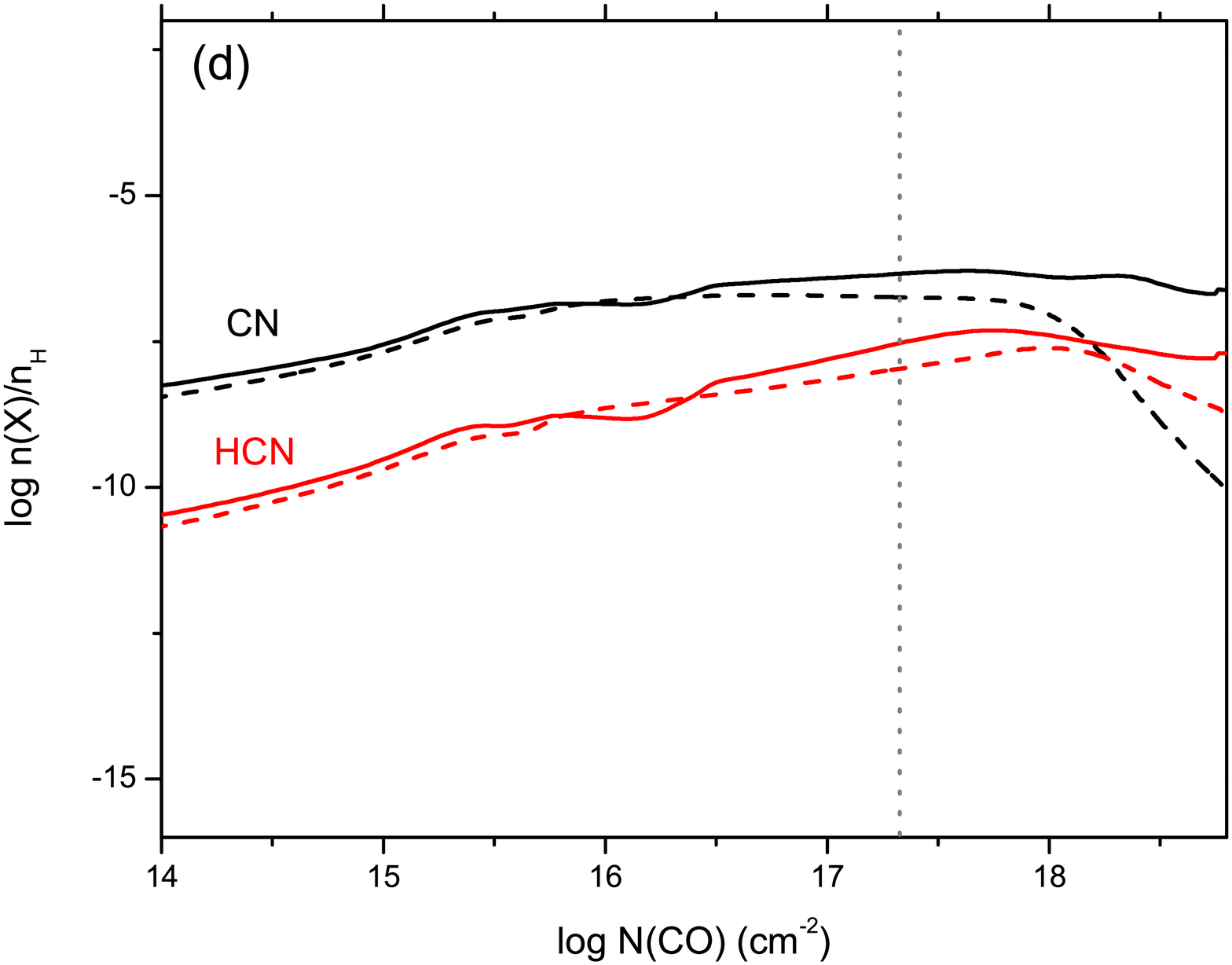}
\end{minipage}

\caption{Relative abundances for a carbon-rich gas (C/O = 1.5, solid
lines) and for an oxygen-rich gas (C/O = 0.66, dashed lines)
for: (a) O-bearing molecules; (b) C-bearing molecules; (c)
S-bearing molecules; (d) CN and HCN.  The vertical line
indicates the boundary between the H$_2$ and CO regions.  The
line at the top indicates the H$_2$ and CO regions in the nebula.}
\label{fig:C_O_models}
\end{figure*}

On the other hand, CO exhausts the less abundant element (C or O) in the CO region, such that the molecular densities in a carbon-rich gas are significantly different from those in an oxygen-rich gas. C-bearing molecules, such as those of the CH$_n$ and C$_n$ families, are highly favored in a carbon-rich gas, while OH, H$_2$O, and NO are abundant in an oxygen-rich gas. The chemistry of second-row elements, such as Si and S, is also only affected by the C/O ratio in the CO region. In the specific case of the species CN, HCN, and HNC, in the CO region their presence is highly favored in a carbon-rich nebula, which does not occur if the gas is oxygen-rich. Figure \ref{fig:C_O_models} illustrates these statements. 

We also studied models with different N/O ratios, since PNe detected in CO and H$_2$ have, in general, high or moderate N/O \citep[e.g.][]{HugginsEtAl1996, KastnerEtAl1996}. Our results show that models with different nitrogen abundances result in similar molecular concentrations. Thus, the N/O ratio correlation with the detection of molecular lines in PNe may be mainly related to other properties of PNe that present moderate or high N/O ratio (for example, high progenitor mass, high central star temperature) rather than the abundance of N by itself.

\section{The importance of X-rays for the molecular chemistry} \label{sec:discussion_x}

In this section we compare our results with observational data from the literature. The results are presented and discussed in terms of column densities, since molecular data obtained from the observations are generally given by these quantities. 

\begin{figure*}[!htb]
\begin{minipage}[t]{0.45\linewidth}
\centering
\includegraphics[width=1 \columnwidth,angle=0]{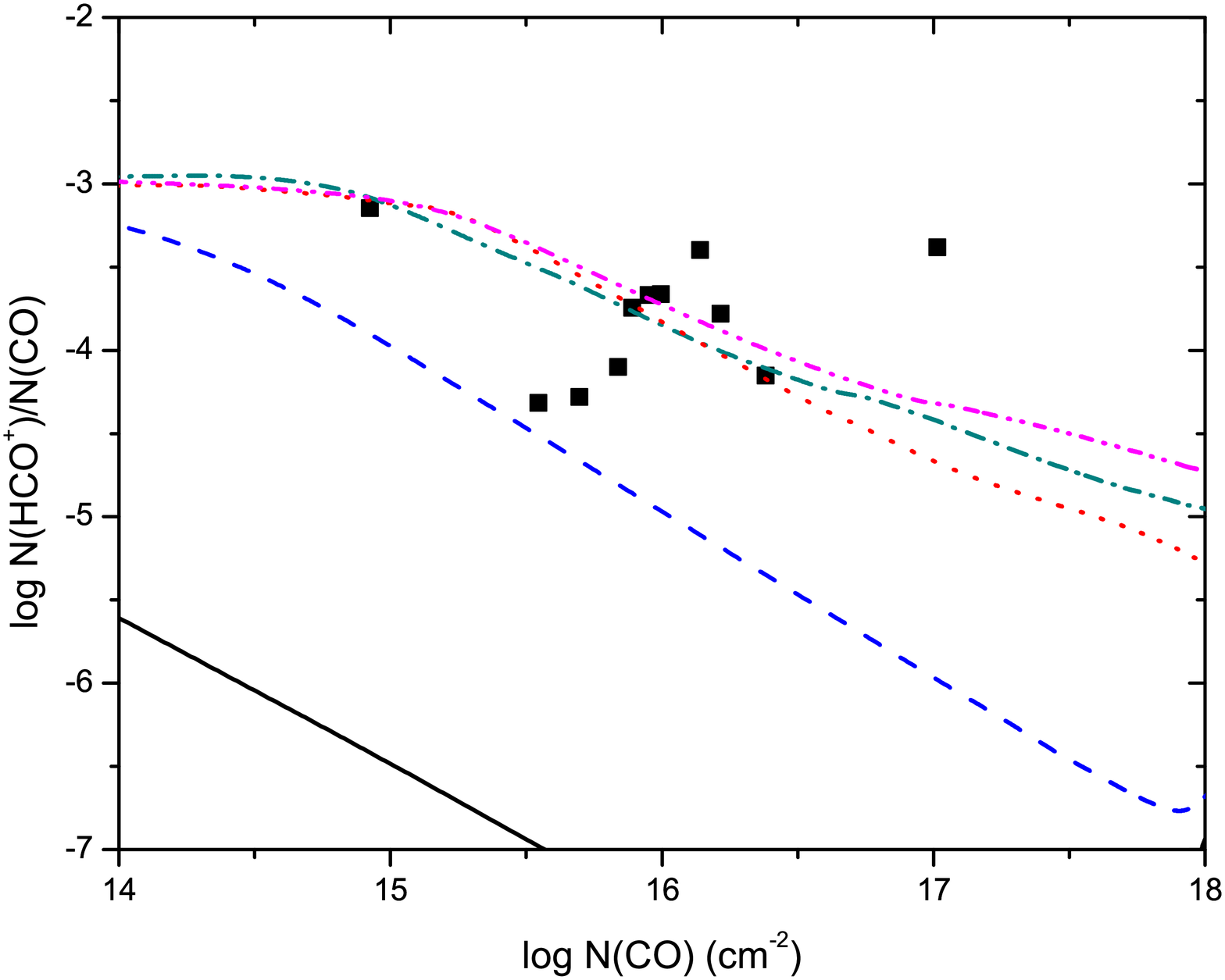}
\end{minipage}
\hspace{0.4cm}
\begin{minipage}[t]{0.45\linewidth}
\centering
\includegraphics[width=1 \columnwidth,angle=0]{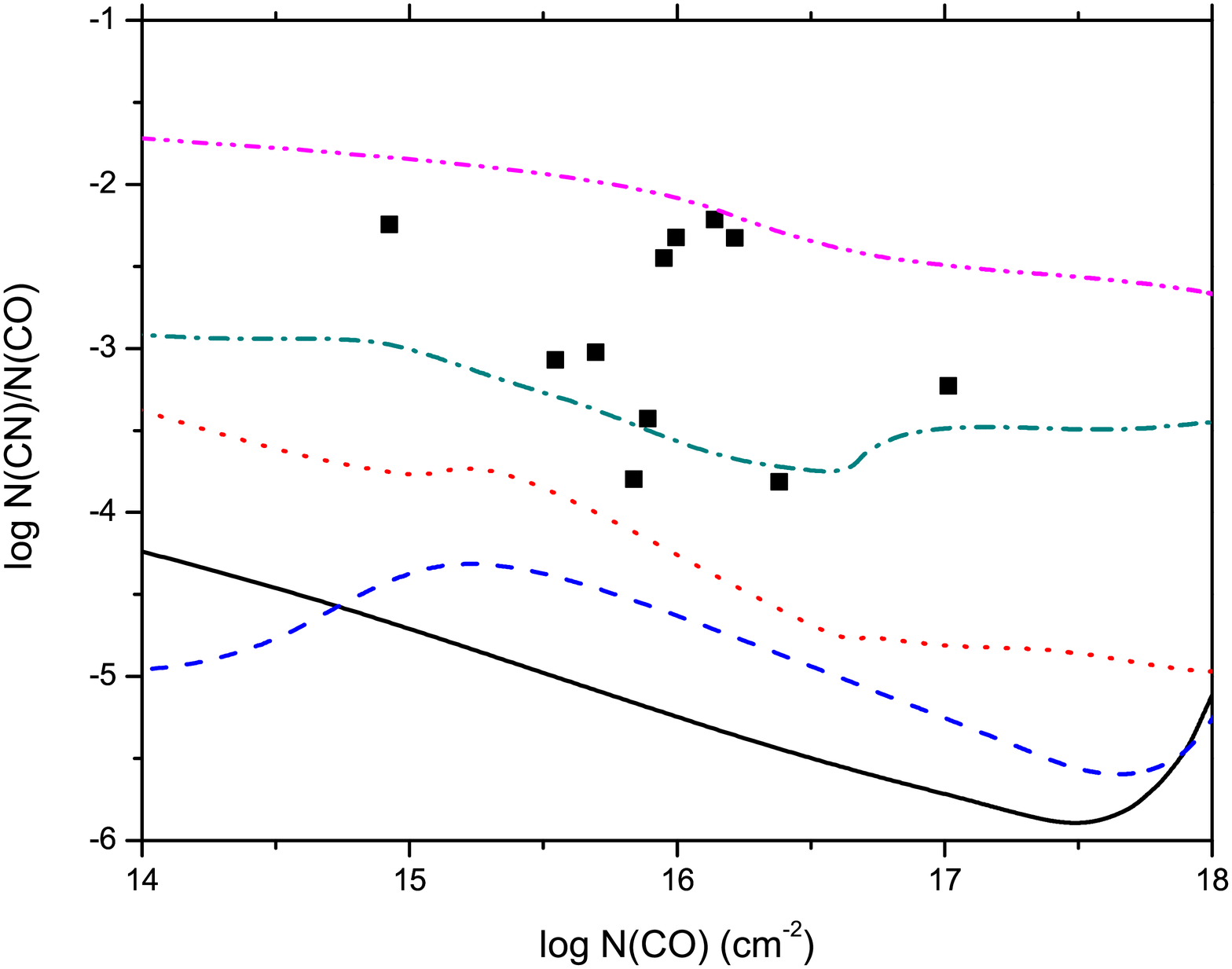}
\end{minipage}

\begin{minipage}[t]{0.45\linewidth}
\centering
\includegraphics[width=1 \columnwidth,angle=0]{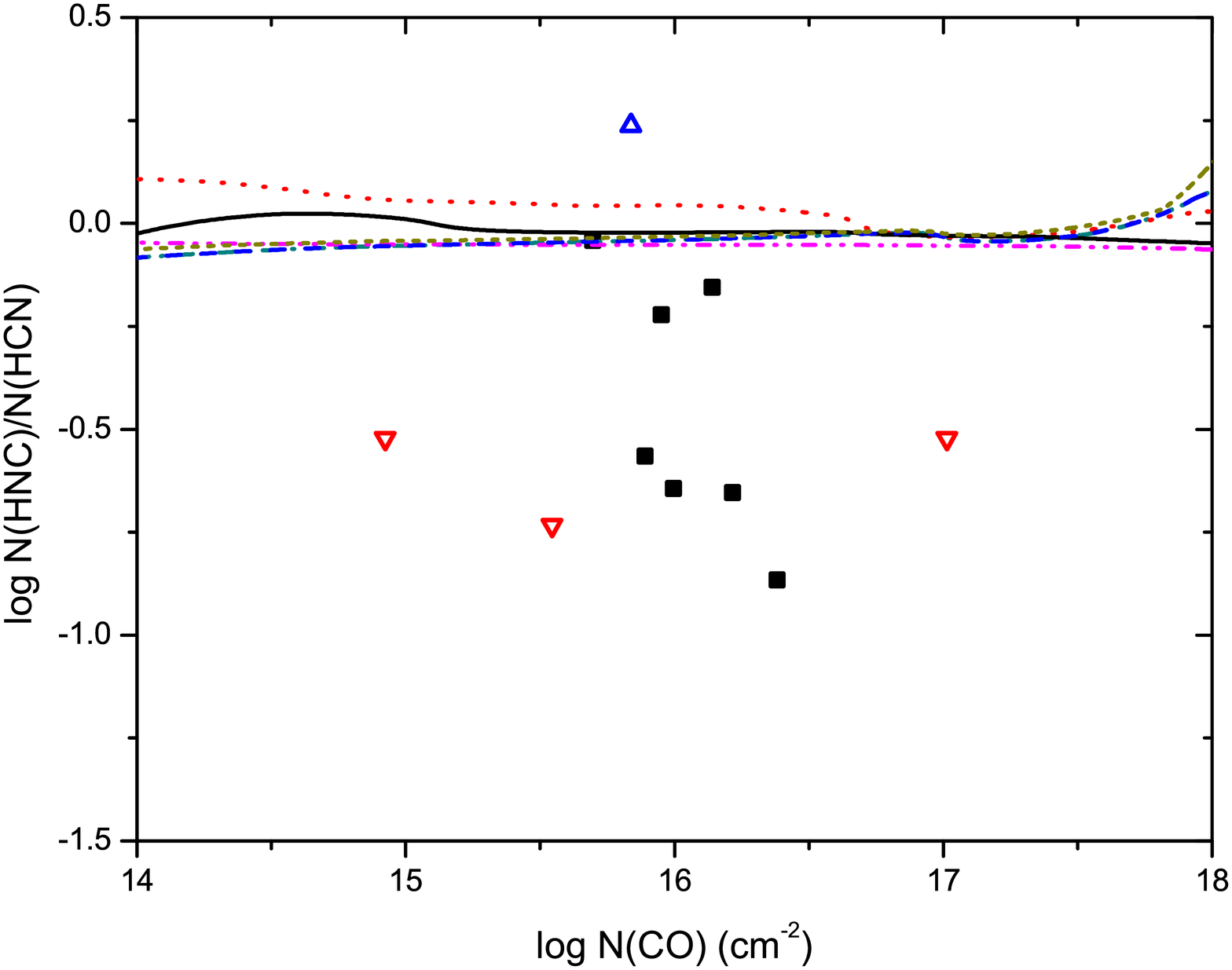}
\end{minipage}
\hspace{0.5cm}
\begin{minipage}[t]{0.45\linewidth}
\centering
\includegraphics[width=1 \columnwidth,angle=0]{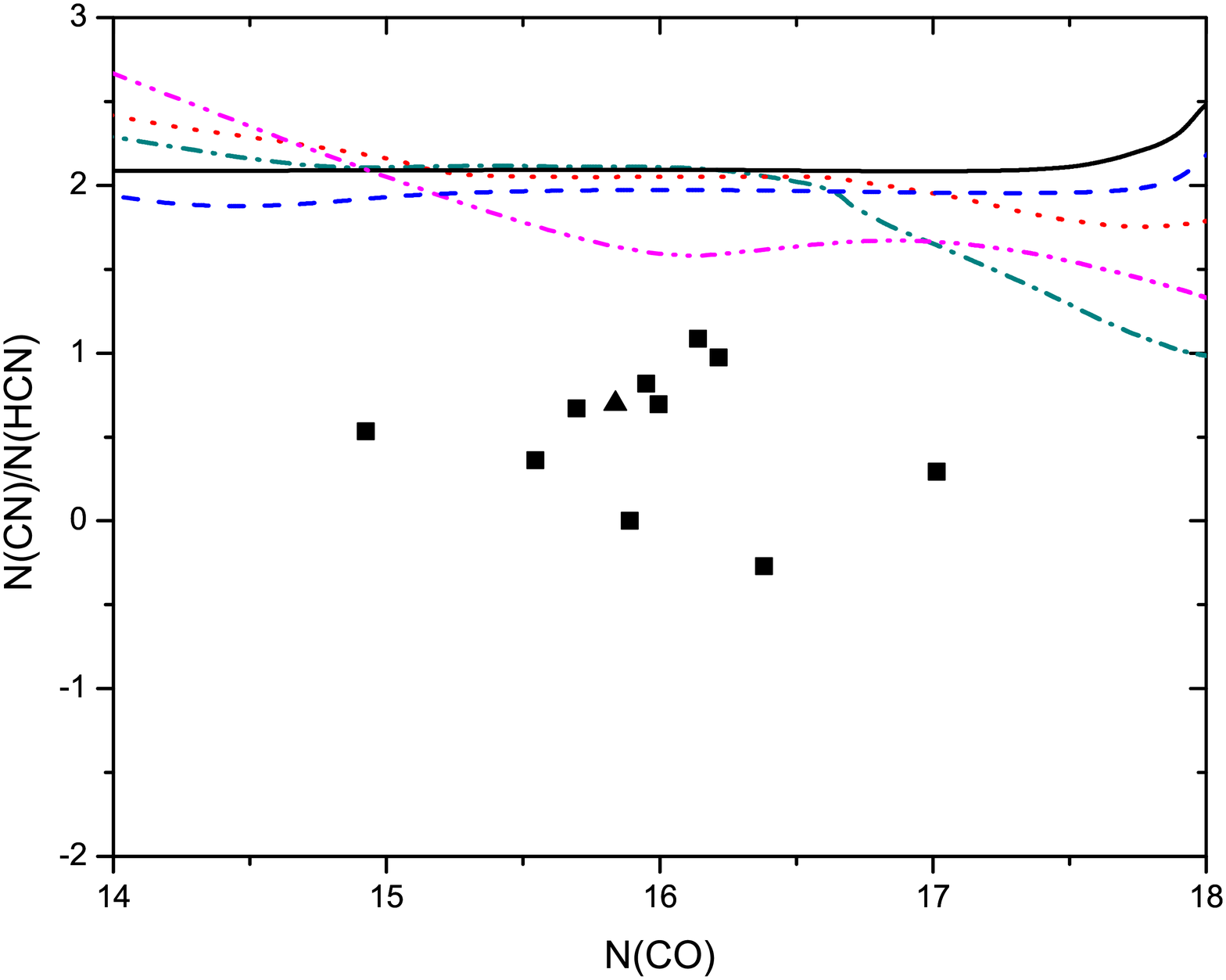}
\end{minipage}

\caption{Column density ratios versus $N($CO$)$. Lines correspond to models
T$_*$ = 5 $\times$ 10$^4$ K and L$_X$ = 0 erg s$^{-1}$ (solid lines);
T$_*$ = 10$^5$ K and L$_X$ = 0 erg s$^{-1}$ (dashed lines);
T$_*$ = 3 $\times$ 10$^5$ K and L$_X$ = 0 erg s$^{-1}$ (dotted lines);
T$_*$ = 10$^5$ K and L$_X$ = $10^{31}$ erg s$^{-1}$ (dot-dashed lines);
T$_*$ = 10$^5$ K and L$_X$ = $10^{32}$ erg s$^-1$ (dot-dot-dashed lines).
Other parameters are those from the standard model. 
Squares and triangles represent observational data, with triangles indicating upper or lower limits, according to their orientation. (A color version of this figure is available in the online journal).}
\label{fig:molecules_relationships}
\end{figure*}

In Fig. \ref{fig:molecules_relationships} we compare column density ratios inferred from observations with our results. Ratios are given as a function of the column density of CO. Symbols represent observational data from the literature \citep{BachillerEtAl1997, JosselinAndBachiller2003} and curves correspond to our models. Two of these models include X-ray emission from a hot bubble, with different values for $L_X$. Two models for which the central star is the only energy source are also shown: one model with a very hot star ($T_* = 3 \times 10^5$ K) and another with a low-temperature star ($T_* = 5 \times 10^4$ K). Other parameters are the same as those from the standard model, except for the dust-to-gas mass ratio which is equal $10^{-2}$. As can be seen in the figure, the $N($HCO$^+)/N($CO$)$ and $N($CN$)/N($CO$)$ ratios are not reproduced by models without significant quantities of X-rays (situation represented by the model with a low-temperature central star).

The molecular chemistry in PNe is sensitive to the stellar temperature mainly because hot stars can emit a significant quantity of X-ray photons ($h\nu \gtrsim 50$ eV). Owing to the low ionization cross-sections for X-rays, these photons can cross long distances in the nebula. As a result, they can increase the ionization rate in regions where the gas is predominantly neutral, favoring ion-molecule reactions. Additionally, X-ray photons can maintain a relatively high electron density in those regions, which can determine molecular density ratios. The importance of the effect of X-rays on the chemistry was stressed by studies of XDRs, an acronym for X-ray dominated regions \citep[e.g.][]{MaloneyEtAl1996}.

As can be seen in Fig. \ref{fig:molecules_relationships}, the model with $T_* = 3 \times 10^5$ K (and $L_X = 0$) shows that for the  $N($HCO$^+)/N($CO$)$ ratio, the data can be explained by models that include a very hot central star, source of significant X-ray emission. In the set of models with $L_X = 0$, the minimum temperature of the central star that could fit the data for $N($HCO$^+)/N($CO$)$ is $T_* \sim 1.5 \times 10^5$ K.

However, notice that even the lowest values of $N($CN$)/N($CO$)$ cannot be reproduced by models that do not include hot bubble X-ray emission. Indeed, if a hot bubble emission is taken into account, at least in the range of values inferred by the observations (10$^{31}$ erg s$^{-1}$ $\lesssim L_X \lesssim$ 10$^{32}$ erg s$^{-1}$), the effects of X-rays emitted by the star are comparatively negligible. 

Theoretical values of the $N($HNC$)/N($HCN$)$ ratio only reproduce the highest values obtained from the observations. Since HCN and HNC have similar chemical paths, with similar rate coefficients, the value of $N($HNC$)/N($HCN$)$ is near unity in most of our models. The lowest observational values are not reproduced by any set of free parameters. In the astrochemical literature, the derived $N($HNC$)/N($HCN$)$ from observations in the interstellar medium ranges from 0.01 to 10 \citep{TurnerEtAl1997,HirotaEtAl1998,StauberEtAl2004}. This range of values is only poorly understood yet. Isomeration, alternative formation routes, or reactions that are processed efficiently for one of those molecules and not for the other in some particular physical condition are some possible explanations proposed in the literature \citep[e.g.][]{GoldsmithEtAl1981, GoldsmithEtAl1986, TurnerEtAl1997,HirotaEtAl1998,StauberEtAl2004}.
 
The high $N($HCN$)/N($CN$)$ ratio compared to the observational data indicates that $N($HCN$)$ is underestimated, since $N($CN$)$ is well-reproduced compared to $N($CO$)$ in our models. The difficulty in reproducing the $N($HCN$)/N($CN$)$ ratio was also found by \citet{HoweEtAl1994}, \citet{AliEtAl2001}, and \citet{RedmanEtAl2003}. \citet{HasegawaEtAl2000} only reproduced the low value of $N($CN$)/N($HCN$)$ for NGC 7027 by postulating a high-temperature gas in the neutral region ($T = 800$ K), where the conversion reaction CN $\stackrel{\mathrm{H}_2}{\rightarrow} $HCN is efficiently processed. In our models, the temperature distribution is consistently calculated assuming thermal equilibrium. Our models do not show neutral regions with such high temperature and, therefore, the reaction above is not efficient. Another heating mechanism (as shock for example) should therefore be present. Another possibility that could explain the low HCN density in the models is the absence of an important reaction in the chemical database \citep[e.g.][]{TurnerEtAl1997,StauberEtAl2004} or inaccurate rate coefficients.

In brief, most of the data inferred from the observations can be explained by X-ray radiation. \citet{AliEtAl2001} also concluded that X-rays dominate the chemistry and lead to the observed molecular composition in the three PNe studied by them. However, while \citet{AliEtAl2001} justified the existence of high concentrations of HCO$^+$ by the presence of H$_3^+$, our results show that in most cases ($N($CO$) \lesssim 10^{17}$ cm$^{-2}$) the main formation processes occur via CO$^+$, whose density is also enhanced by X-ray radiation. Moreover, \citet{AliEtAl2001} justified the inclusion of X-ray emission in their models because of the high temperature of the central stars. Although the X-ray emission from a hot star is a possible explanation for $N($HCO$^+)/N($CO$)$ ratio, our study shows that the $N($CN$)/N($CO$)$ ratio is only adequately reproduced by models including X-ray emission from a hot bubble. Moreover, stellar temperatures would need to be atypically high to reproduce the $N($HCO$^+)/N($CO$)$ ratio inferred from the observations.

\begin{figure}[!htb]
\begin{center}
\includegraphics[width=1 \columnwidth,angle=0]{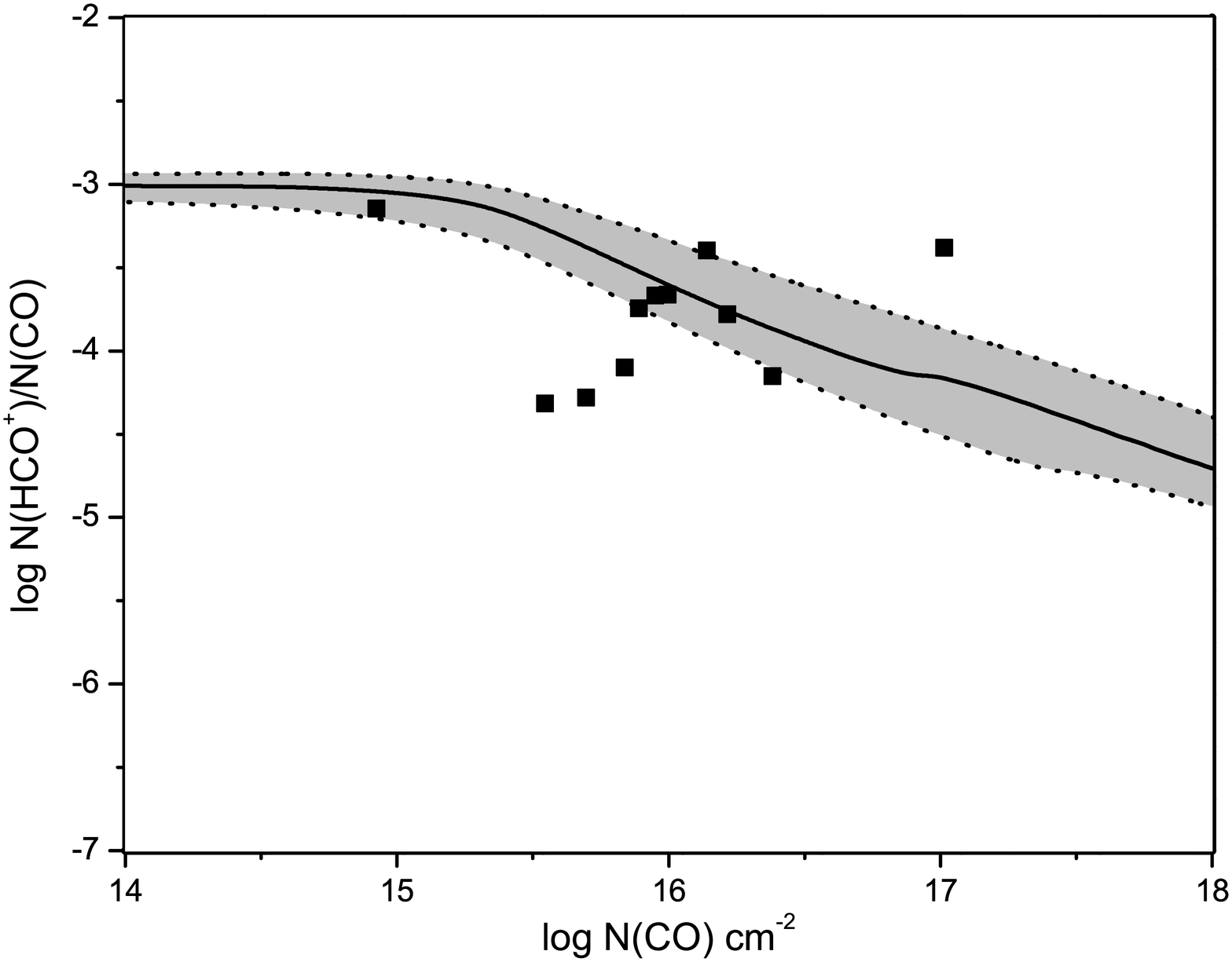}
\includegraphics[width=1 \columnwidth,angle=0]{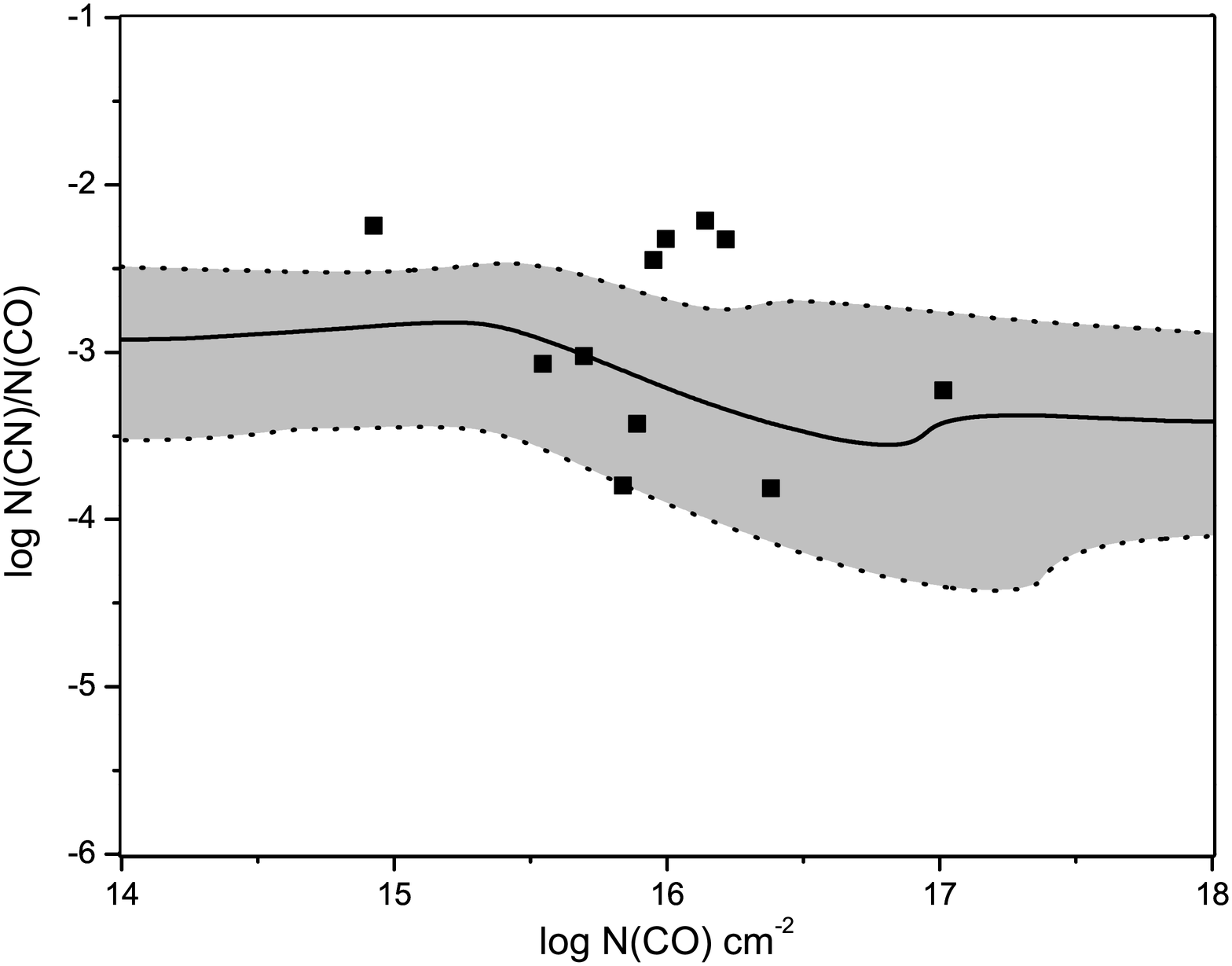}
\caption{$N($HCO$^+)/N($CO$)$ (top panel) and $N($CN$)/N($CO$)$ (bottom panel) 
versus $N($CO$)$. The gray area indicates the range of values obtained with our grid of models with $L_*$, $n_H$ and $M_d/M_g$ in the range given in Table \ref{tab:FreeParameters}, and $L_X$ and $T_*$ the same for the standard model. The solid curve represents the standard model. Squares represent observational data.}
\label{fig:Rgg}
\end{center}
\end{figure}

As said above, the curves presented in Fig. \ref{fig:molecules_relationships} correspond to models with input parameters of the standard model, except those for the stellar temperature and for the X-ray flux. Some deviation is expected for different sets of parameters. As an example, in Fig. \ref{fig:Rgg}, the $N($HCO$^+)/N($CO$)$ and $N($CN$)/N($CO$)$ ratios are presented for models with $L_*$, $n_H$ and $M_d/M_g$ in the range given in Table \ref{tab:FreeParameters}, and $L_X$ and $T_*$ the same as for the standard model. The parameter for which the molecular chemistry is more sensitive, besides the X-rays discussed above, is the dust-to-gas mass ratio (the highest and lowest curves in the figure for both ratios are for models with $M_d/M_g = 10^{-2}$ and $M_d/M_g = 10^{-3}$, respectively). The presence of dust affects the molecular chemistry in three ways: (1) through absorption of the radiation, (2) because they act as catalysis for the formation of H$_2$ molecules, and (3) through the photoelectric effect, which heats the gas. As a consequence, for higher dust-to-gas ratios the H$_2$ region is wider, the H$^0$ region is narrower, and both regions are closer to the radiation central source. Consequently, the density of ionized species is higher and the ion-molecule reactions are more effective. Although the effect of dust on molecular chemistry is significant, it is not possible to reproduce most of the column density ratios inferred from the observations without including a source of X-rays.

The molecular composition, as well as the physical conditions in each region (gas temperature, ionic distribution), in a plot $N($X$)/N($Y$)$ versus $N($CO$)$ is weakly dependent on the gas density and on the stellar luminosity. Accordingly, different density profiles do not change the discussion about the ratios between molecular densities. The only exception is the ratio $N($CN$)/N($CO$)$. Since the concentration of these two molecules are not strongly connected, small differences in gas temperature, ionic abundances, and incident flux convert in non-negligible differences in the referred ratio, mainly in the H$_2$ region. We emphasize that this ratio is sensitive to $n_H$ and $L_*$, although the effects of the other parameters (X-ray luminosity, star temperature and dust-to-gas mass ratio) discussed above are more significant. In general, for higher gas density (or lower stellar luminosity), the ratio $N($CN$)/N($CO$)$ is higher.

\section{The molecular mass and the missing mass problem in PNe} \label{sec:molecular_mass}

Masses of PNe progenitors in the main-sequence can be as high as 8 M$_\odot$ as indicated by the detection of white dwarfs in open clusters \citep[e.g.][]{ KaliraiEtAl2007,KaliraiEtAl2008, WilliamsEtAl2009}. On the other hand, masses obtained for the central stars of PNe and for the nebular component can be as high as 1.2 and 0.3 M$_\odot$, respectively \citep[e.g.][]{Kwok2000}. Consequently, there must be some mass that is not being accounted for. The difference between the total mass of a PNe and that of the main-sequence progenitor is known as the missing mass problem in PNe \citep[e.g.][]{Kwok1994}. Because the above calculation of the nebular masses only takes into account the ionized mass, \citet{Kwok1994} suggests that the missing mass problem could be solved if the neutral matter in PNe is taken into account. 

A more recent determination of the total mass associated with PNe was obtained by \citet{BernardSalasAndTielens2005}, who calculated for two objects the sum of the nebular mass (ionized, atomic, and molecular) and the mass of the central star. Using molecular masses derived from CO observations by \citet{HugginsEtAl1996}, \citet{BernardSalasAndTielens2005} found good agreement between the mass expected for the progenitor star of NGC 6302 \citep[$\sim 4.5$ M$_\odot$,][]{MarigoEtAl2003} and the total mass estimated for this PN ($\sim 3.9$ M$_\odot$). On the other hand, for NGC 7027 part of the ejected mass seems to be missing ($\sim 2-3$ M$_\odot$), even taking the neutral matter into account. It must be noted that for the former PN the atomic mass corresponds to almost the whole nebular mass, while for the second one most of the gas is molecular.

The molecular masses derived by \citet{HugginsEtAl1996} are obtained from the intensity of the rotational line 2-1 of the CO molecule and approximations for the ratio between the CO and H$_2$ column densities. From our models, however, we conclude that the method adopted by \citet{HugginsEtAl1996} can underestimate the molecular content of PNe, as is discussed below.

The molecular mass was obtained by \citet{HugginsEtAl1996} from the rotational line 2-1 of CO molecule according to the expression

\begin{equation}
M_{\textrm{\tiny mol}} = 2.6 \times 10^{-10}FD^2 / f,
\label{eq:M_mol_Huggins}
\end{equation}
\noindent where $F$ is the CO line flux (in K km s$^{-1}$ arcsec$^2$), $D$ is the distance to the PN (in kpc), $f$ is the column density of CO relative to that of hydrogen, and $M_{\textrm{\tiny mol}}$ is given in solar masses. The formula includes a correction for the presence of helium with a representative abundance value of He/H = 0.1.

To obtain a value for $f$, \citet{HugginsEtAl1996} assumed the following approximations: (1) all hydrogen nuclei are in the molecular form; (2) CO is fully associated. With these simplifications the authors adopted a value for $f$ equal to 3 $\times 10^{-4}$, based on the C/H and O/H ratios in PNe where both abundances have been measured. Then, for $f$

\begin{equation}
f = \frac{N(\mathrm{CO})}{N(\mathrm{H})} = \frac{N(\mathrm{CO})}{2 N(\mathrm{H}_2)} = 3 \times 10^{-4},
\label{eq:Huggins_f}
\end{equation}
\noindent 

Our models, however, show that CO is fully associated only in the CO region (see Fig. \ref{fig:SM_abundances}), which is mostly dissociated in the H$_2$ region. As a consequence, the ratio between integrated quantities, such as column densities, depends on the specific nebula, i. e., on its total mass and how the molecules are distributed along the nebula. As stated above, the distribution of the molecules depends, in turn, on the incident radiation spectrum and on the nebula parameters (density, elemental abundance, etc.). 

\begin{figure}[!htb]
\begin{center}
\includegraphics[width=1 \columnwidth,angle=0]{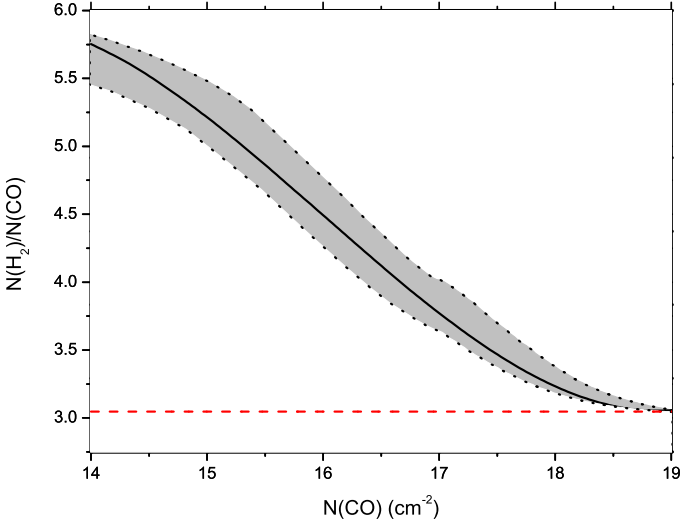}
\caption{Column density ratio $N($H$_2)/N($CO$)$. The gray area indicates the range of values obtained with our grid of models. The solid line represents the values obtained for the standard model (Eq. \ref{eq:YH2_CO}). The horizontal dashed
line corresponds to the hypothesis of \citet{HugginsEtAl1996} that CO is fully associated, applied to the chemical composition of the standard model.}
\label{fig:NH2_NCO_relation}
\end{center}
\end{figure}

Figure \ref{fig:NH2_NCO_relation} shows the $N($H$_2)/N($CO$)$ ratio (= $0.5f^{-1}$) for a set of models that spans the range of physical parameters given in Table \ref{tab:FreeParameters} (excepting models with a chemical composition different from that adopted for the standard model and models with $L_X = 0$). 

Based on our models we can represent the $N($H$_2)/N($CO$)$ ratio according to the equation

\begin{eqnarray}
 \nonumber \log \frac{N(\mathrm{H}_2)}{N(\mathrm{CO})}  & = & -107.5 + 22.50 Y_{CO} - 1.453 Y_{CO}^2 \\
 &  & + 3.028 \times 10^{-2} Y_{CO}^3,
 \label{eq:YH2_CO}
\end{eqnarray}
\noindent where $Y_{CO} = \log N($CO$)$. This relation is valid for $10^{14} \leq N($CO$) \leq 10^{19}$ cm$^{-2}$ and is derived for the standard model. It corresponds to the solid curve in Fig. \ref{fig:NH2_NCO_relation}. For the other models shown in this figure the maximum deviation is of a factor two. For models with L$_X$ = 0, not shown in the figure, the results can differ by up to a factor five.

Taking into account the results from models with values for the C and/or O abundance different from that of the standard model, the $N($H$_2)/N($CO$)$ ratio derived from Eq. \ref{eq:YH2_CO} must be multiplied by the following correction factor

\[
\frac{4.79 \times 10^{-4}}{ X_{C,O}/H},
\]
\noindent where $X_{C,O}/H$ is the abundance of the less abundant element between C and O.

As said above, \citet{HugginsEtAl1996} assumed that CO is fully associated. The horizontal dashed line in Fig. \ref{fig:NH2_NCO_relation} corresponds to this hypothesis applied to the chemical composition of the standard model. Note that the assumption is only correct for very high values of the CO column density, as can be seen in Fig. \ref{fig:NH2_NCO_relation}.

\begin{figure}[!htb]
\begin{center}
\includegraphics[width=1 \columnwidth,angle=0]{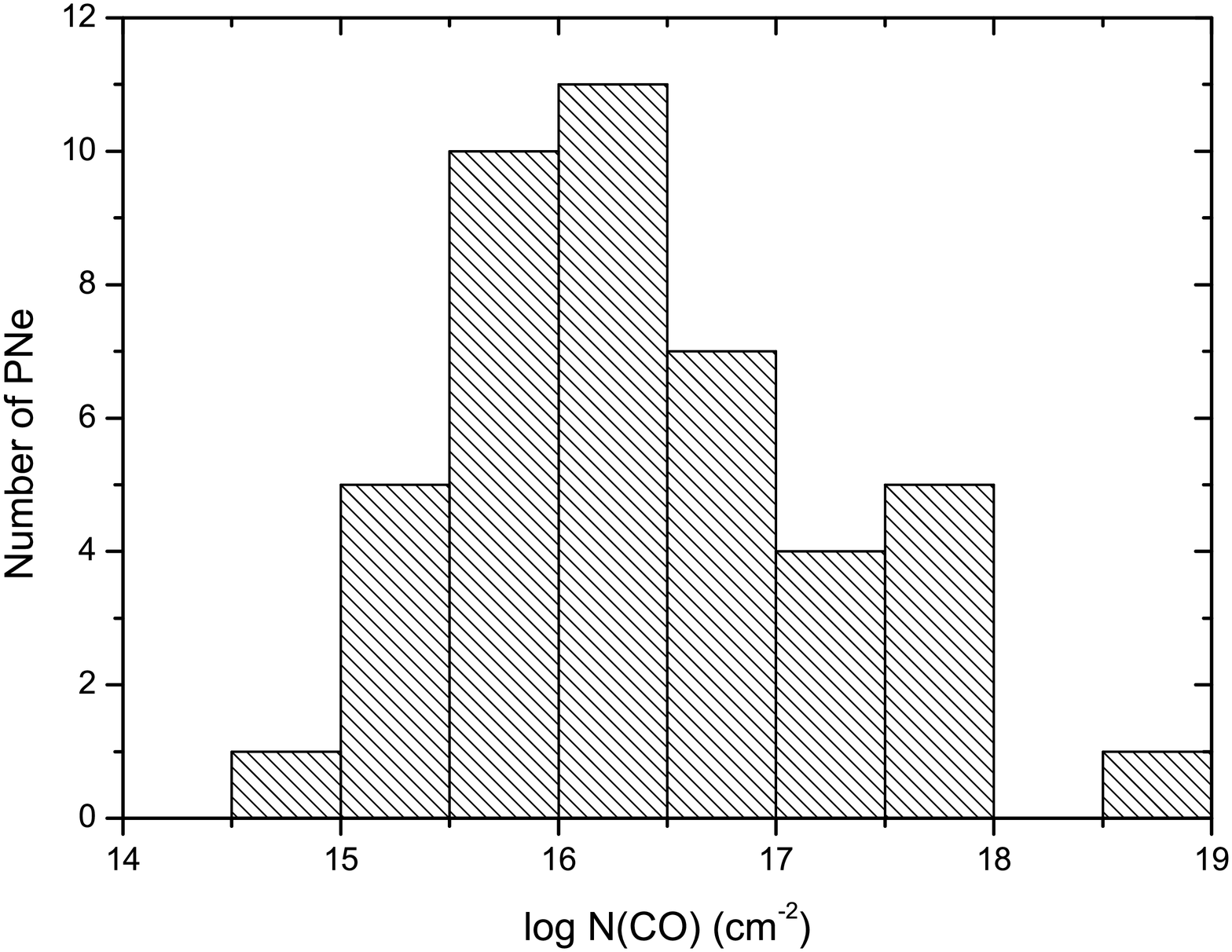}
\caption{Distribution of $N($CO$)$ for PNe. Data from \citet{HugginsEtAl1996}.}
\label{fig:NCO_counting}
\end{center}
\end{figure}

The distribution of PNe with CO column densities calculated by Huggins et al. (1996) is shown in Fig. \ref{fig:NCO_counting}. Comparing this histogram with Fig. \ref{fig:NH2_NCO_relation}, it can be noticed that there is a large number of PNe in their sample for which their approximation for the ratio $N($H$_2)/N($CO$)$ is not valid. For these objects the obtained molecular mass is then underestimated.

It is important to keep in mind that when \citet{HugginsEtAl1996} determined 
the column density, they did not take into account individual characteristics 
of each nebula, such as geometry and angular size. That is, they assumed a 
simplified correction factor to take into account the beam dilution for 
objects with diameters smaller than the beam. In these cases, the 
observational CO column density can differ from the radial column density, as 
obtained from the models.

\begin{figure}[!htb]
\begin{center}
\includegraphics[width=1 \columnwidth,angle=0]{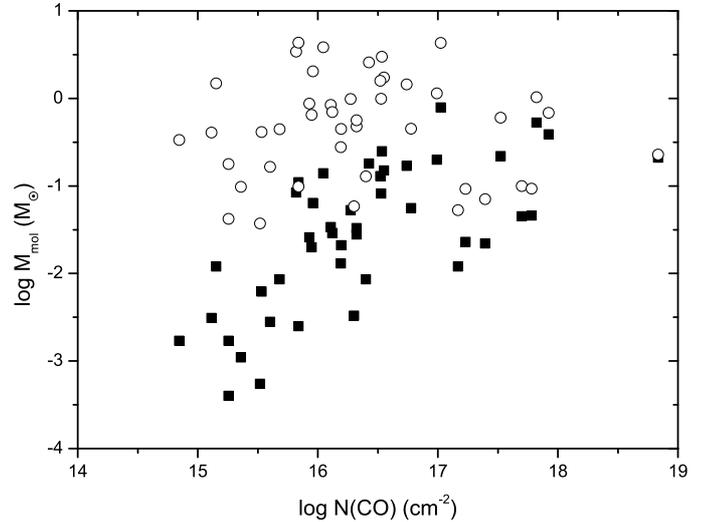}
\caption{Molecular mass versus the CO column density. Squares show the masses obtained by \citet{HugginsEtAl1996}, and circles indicate masses obtained by the method described in the text.}
\label{fig:Mass_corrected}
\end{center}
\end{figure}

The molecular masses obtained by \citet{HugginsEtAl1996} for their sample can then be recalculated from Eq. \ref{eq:M_mol_Huggins}, assuming the $N($H$_2)/N($CO$)$ ratio (= $0.5f^{-1}$) from Eq. \ref{eq:YH2_CO} scaled to $X_{C,O}/H = 3 \times 10^{-4}$ implicitly assumed by them. The new values for the masses and those obtained by \citet{HugginsEtAl1996} are compared in Fig. \ref{fig:Mass_corrected}. As expected, the differences between new and previous masses are greater for nebulae with lower values for $N($CO$)$. Our results do not show the trend of increasing molecular mass with increasing $N($CO$)$ noticed by \citet{HugginsEtAl1996}. It must be remarked that $N($CO$)$ is not necessarily related to the nebular size and mass, but depends on the distribution of the CO molecule on the nebula which, as said above, depends on the incident radiation and on the absorption properties of the nebula. 

According to observations and evolutionary models, masses for the central star of PNe are mostly in the range 0.55 M$_\odot$ to 0.65 M$_\odot$ \citep[e.g.][]{StasinskaEtAl1997}, which corresponds to progenitor stars with masses approximately in the range 1 M$_\odot$ to 3 M$_\odot$ \citep[e.g.][]{Kwok2000}.  Thus, the total nebular mass (given by the difference between the mass of the progenitor star and that of the central star of the planetary nebulae) should be in the range 0.35 M$_\odot$ to 2.45 M$_\odot$. 

\begin{figure}[!htb]
\begin{center}
\includegraphics[width=1 \columnwidth,angle=0]{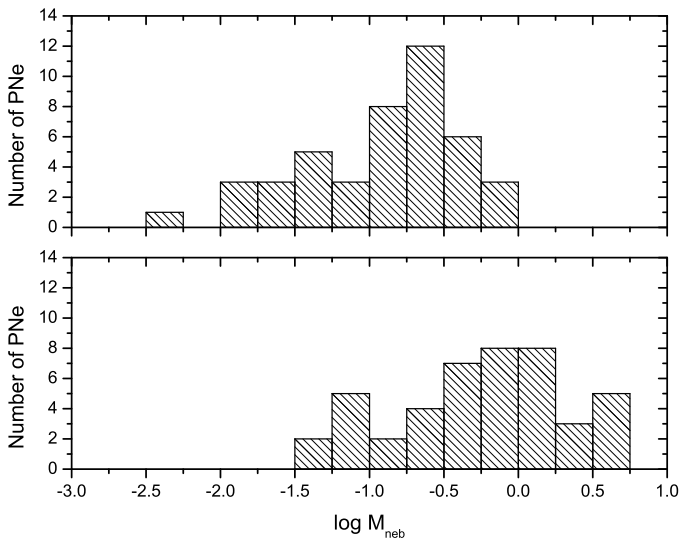}
\caption{Nebular mass distribution for the \citep{HugginsEtAl1996} sample: their calculation (top panel), and after recalculation of the molecular mass according to the method described in the text (botton).}
\label{fig:Hist_Nebular_Mass}
\end{center}
\end{figure}

From masses for the ionized gas derived from radio observations and their calculated molecular masses, \citet{HugginsEtAl1996} obtained values for nebular masses in the range $10^{-3}$ to 1 M$_\odot$. These results do not reproduce the range given above for the expected nebular masses of PNe.
The distribution of nebular masses given by \citet{HugginsEtAl1996} is shown in the upper panel of Fig. \ref{fig:Hist_Nebular_Mass}.  

The nebular mass can also be calculated from our models. However, to obtain the total mass for the nebulae of the \citet{HugginsEtAl1996} sample, a specific model for each nebula should be obtained. This is beyond the scope of this paper since we merely aimed for an estimate of the nebular mass, after the correction of molecular masses proposed in the present paper. Summing the ionized mass adopted by \citet{HugginsEtAl1996} and the corresponding molecular mass recalculated as shown above for each planetary of their sample, the obtained nebular mass distribution is that shown in the lower panel of Fig. \ref{fig:Hist_Nebular_Mass}. The new calculated masses are up to 4.5 M$_\odot$, which agrees better with the expected distribution of nebular masses. Note that this is still a lower limit, since the atomic mass is not taken into account. 

\section{Summary} \label{sec:conclusion}

A self-consistent numerical code was adapted to simulate gaseous nebulae around ionizing stars, from the hot and ionized region to the neutral and cold gas. A grid of models was obtained for stellar and nebular parameters typical of planetary nebulae. 

The spatial distribution of molecules and the main mechanisms of molecular formation and destruction were discussed. The effect of the input parameters on the chemical composition was also analyzed. 

Our results show that a strong X-ray radiation field is needed to explain most of the column density ratios inferred from the observations for species such as CO, CN, and HCO$^+$. The molecular chemistry is also sensitive to the dust-to-gas mass ratio, while the dependence on $L_*$ and $n_H$ is comparatively weak. 

Column density ratios involving HCN and HNC are, however, not reproduced by the models. This problem is also found in other models of the literature. The HCN molecule is underestimated in our models. We suggest that molecular data related to the reactions should be reviewed. An additional heating mechanism in the neutral region could be an alternative explanation.

We showed that the $N($H$_2)/N($CO$)$ column density ratio depends on the value 
of $N($CO$)$. This is due to the distribution of both molecules inside the nebula. Using the relation obtained in the present work, we recalculated the molecular masses previously obtained in the literature, showing that these masses are usually highly underestimated. As a result, we conclude that the missing mass problem can be solved by a more accurate estimate of $N($H$_2$)/$N($CO$)$.

\section*{Acknowledgments}

R.K. acknowledged the financial support from FAPESP Brazil grant 
number 03082/07.

I.A is thankful for the financial support of FAPESP (fellowship 2007/04498-2), CNPq (PDE 201950/2008-1), and CAPES/PRO-DOC. 

We also acknowledge the anonymous referee for the valuable suggestions that helped to improve this paper.


\bibliographystyle{aa}  
\bibliography{bibliografia}			

\label{lastpage}

\end{document}